\documentclass[journal=nalefd, manuscript=letter, layout=twocolumn]{achemso}

\usepackage{amssymb, physics, titlesec}
\usepackage[version=4]{mhchem}
\usepackage[separate-uncertainty = true]{siunitx}
\usepackage[hidelinks]{hyperref}
\usepackage{xcolor}
\usepackage[normalem]{ulem}

\makeatletter
\let\l@addto@macro\relax
\makeatother
\usepackage[fontsize=10pt]{scrextend}
\mciteErrorOnUnknownfalse

\titleformat{\subsection}[runin]{\bfseries}{}{}{}[]
\let\oldmaketitle\maketitle
\let\maketitle\relax

\newcommand{\supp}[1]{Supporting Section~S#1}

\title{Resolving the controversy in biexciton binding energy of cesium lead halide perovskite nanocrystals through heralded single-particle spectroscopy}

\author{Gur Lubin}
\affiliation{Deptartment of Physics of Complex Systems, Weizmann Institute of Science, Rehovot, Israel}

\author{Gili Yaniv}
\author{Miri Kazes}
\affiliation{Deptartment of Molecular Chemistry and Materials Science, Weizmann Institute of Science, Rehovot, Israel}

\author{Arin Can Ulku}
\author{Ivan Michel Antolovic}
\author{Samuel Burri}
\affiliation{School of Engineering, École polytechnique fédérale de Lausanne (EPFL), Neuchâtel, Switzerland}

\author{Claudio Bruschini}
\author{Edoardo Charbon}
\affiliation{School of Engineering, École polytechnique fédérale de Lausanne (EPFL), Neuchâtel, Switzerland}

\author{Venkata Jayasurya Yallapragada}
\affiliation{Deptartment of Physics of Complex Systems, Weizmann Institute of Science, Rehovot, Israel}
\email{venkata-jayasurya.yallapragada@weizmann.ac.il}

\author{Dan Oron}
\affiliation{Deptartment of Molecular Chemistry and Materials Science, Weizmann Institute of Science, Rehovot, Israel}
\email{dan.oron@weizmann.ac.il}

\begin{document}
\twocolumn[
\begin{@twocolumnfalse}
\oldmaketitle
\begin{abstract}
Understanding exciton-exciton interaction in multiply-excited nanocrystals is crucial to their utilization as functional materials. Yet, for lead halide perovskite nanocrystals, which are promising candidates for nanocrystal-based technologies, numerous contradicting values have been reported for the strength and sign of their exciton-exciton interaction. In this work we unambiguously determine the biexciton binding energy in single cesium lead halide perovskite nanocrystals at room temperature. This is enabled by the recently introduced SPAD array spectrometer, capable of temporally isolating biexciton-exciton emission cascades while retaining spectral resolution. We demonstrate that \ce{CsPbBr3} nanocrystals feature an attractive exciton-exciton interaction, with a mean biexciton binding energy of \SI{10}{meV}. For \ce{CsPbI3} nanocrystals we observe a mean biexciton binding energy that is close to zero, and individual nanocrystals show either weakly attractive or weakly repulsive exciton-exciton interaction. We further show that within ensembles of both materials, single-nanocrystal biexciton binding energies are correlated with the degree of charge-carrier confinement.
\end{abstract}
\end{@twocolumnfalse}
]

% -----------------------------------------------------------------------------
% --------------------------------- Figures -----------------------------------
% -----------------------------------------------------------------------------
\begin{figure}[t]
    \centering
    \includegraphics[width=\linewidth]{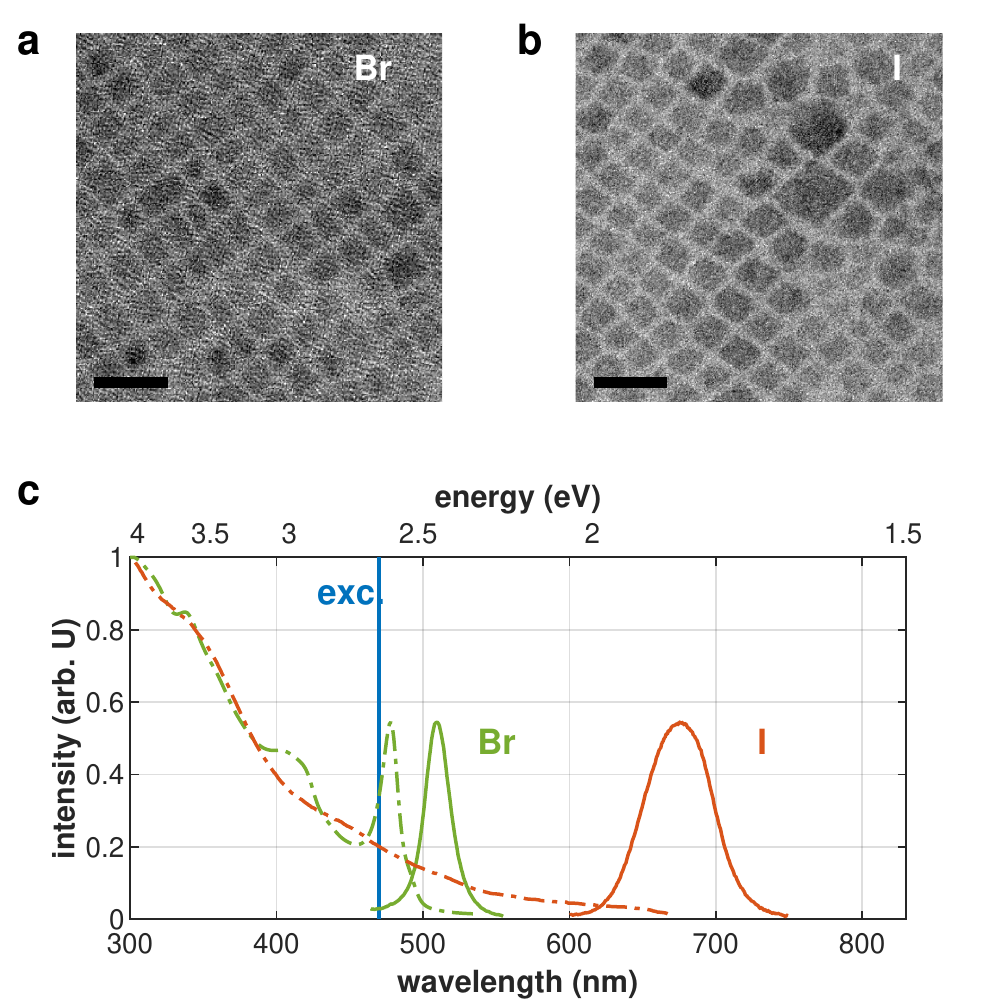}
    \caption{
    \textbf{Particles investigated in this work.} 
    \textbf{a)} Transmission electron micrograph of the \ce{CsPbBr3} NCs investigated in this work. 
    \textbf{b)} Transmission electron micrograph of the \ce{CsPbI3} NCs investigated in this work. Both scale bars are \SI{20}{nm}.
    \textbf{c)} Ensemble emission (solid lines) and absorption (dashed dotted lines) of the \ce{CsPbBr3} (green) and \ce{CsPbI3} (red) NCs. Blue line marks the excitation wavelength (\SI{470}{nm}).
    }
    \label{fig:particles}
\end{figure}

\begin{figure*}[t]
    \centering
    \includegraphics[width=.9\linewidth]{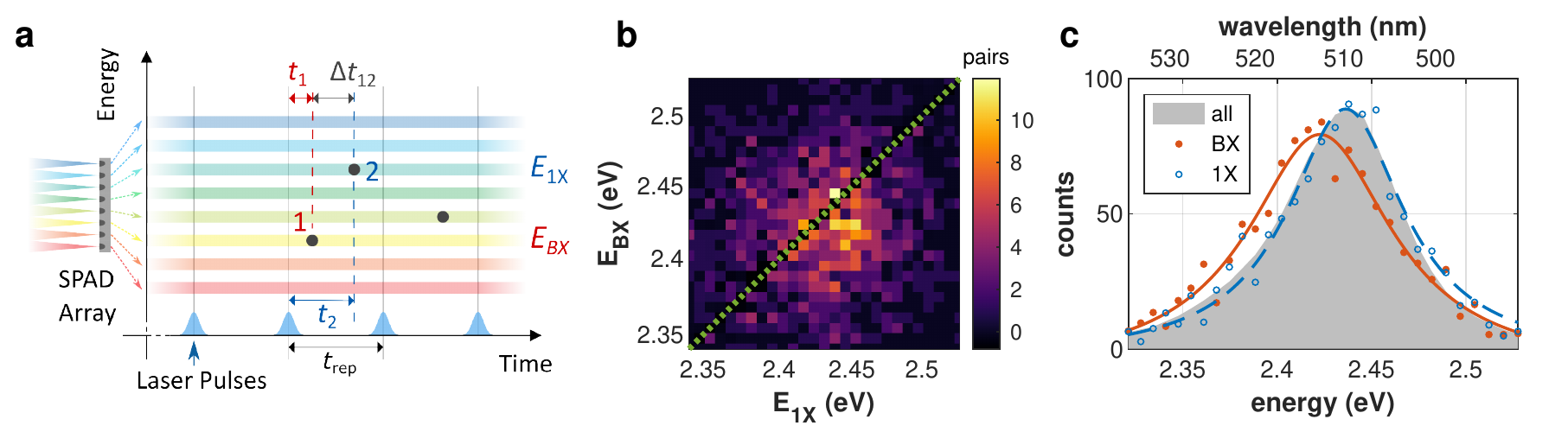}
    \caption{
    \textbf{Heralded spectroscopy of a single particle.} 
    \textbf{a)} A schematic illustration of the heralded spectroscopy scheme. A linear SPAD array is placed at the output of a grating spectrometer such that each SPAD pixel detects a different wavelength. The data from each SPAD pixel consists of the absolute arrival times of photons. By identifying the first and second arriving photons in each coincidence detection (BX and 1X, respectively), their corresponding energies can be extratced ($E_{BX}$ and $E_{1X}$). 
    \textbf{b)} 2D histogram of photon pairs following the same excitation pulse, from a 5-minute measurement of a single \ce{CsPbBr3} NC. Green dashed line is a guide to the eye marking both photons with the same energy (undetectable by the system).
    \textbf{c)} BX spectrum (red dots) and 1X spectrum (blue circles) extracted by full horizontal and full vertical binning of panel (b), respectively. Grey area is the 1X spectrum (normalized) extracted by summing over all detected photons. Red solid line and blue dashed line represent fit of the BX and 1X spectra, respectively, to Cauchy-Lorentz distributions. Binding energy for this specific NC, estimated as the difference between the spectral peaks of the two fits, is $\varepsilon_b = 13.5\pm\SI{1.8}{meV}$.
    }
    \label{fig:single}
\end{figure*}

\begin{figure}[t]
    \centering
    \includegraphics[width=.9\linewidth]{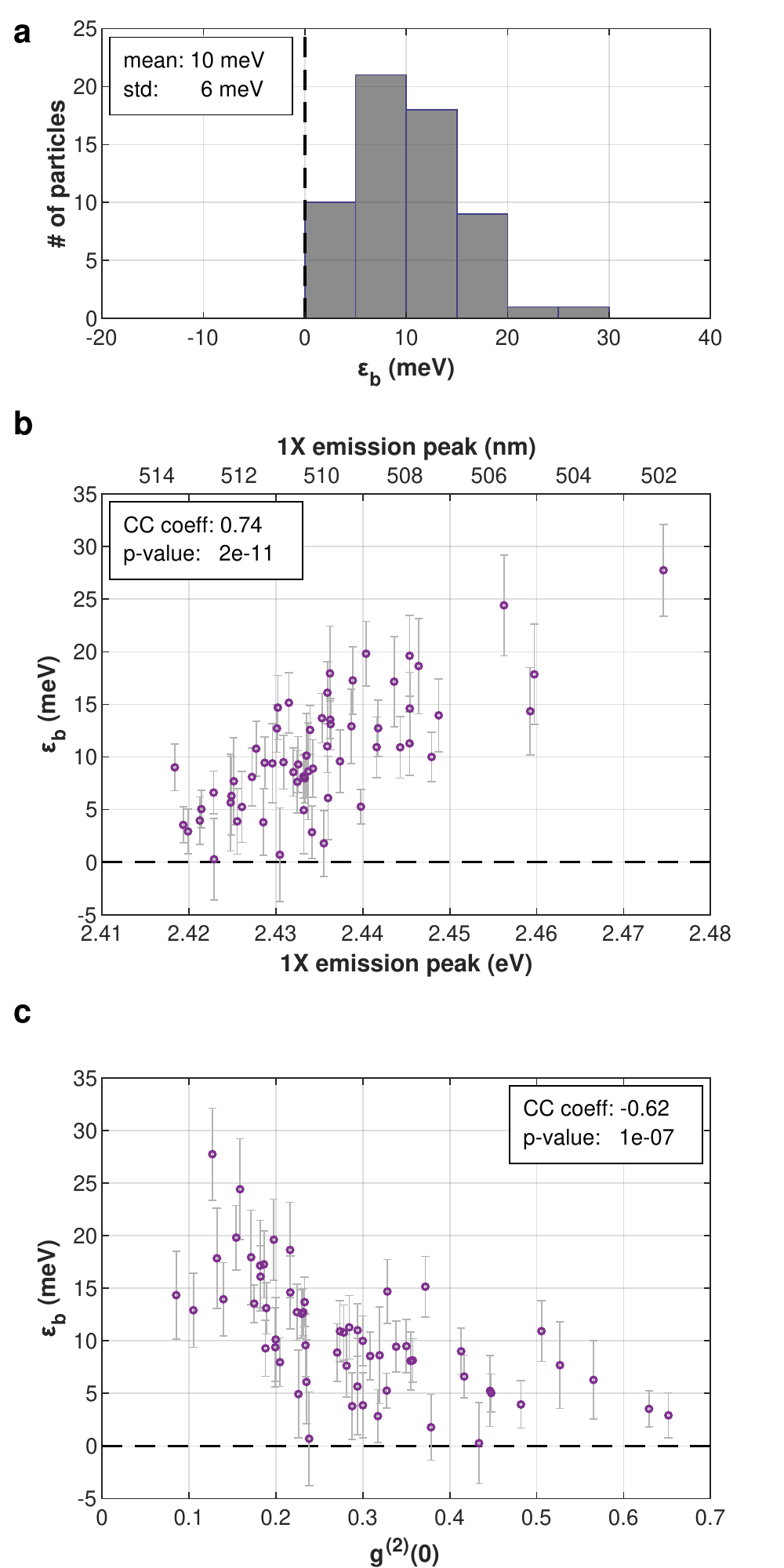}
    \caption{
    \textbf{\ce{CsPbBr3} binding energy.} 
    \textbf{a)} Binding energy histogram for 60 NCs. Mean single-particle error is $\pm\SI{3.1}{meV}$. 
    \textbf{b)} Binding energy as a function of 1X emission peak. 
    \textbf{c)} Binding energy as a function of $g^{(2)}(0)$. 
    \textbf{std} - standard deviation, \textbf{CC coeff} - cross-correlation coefficient. \textbf{p-value} - p-value of the cross-correlation.
    }
    \label{fig:br}
\end{figure}

\begin{figure}[t]
    \centering
    \includegraphics[width=.9\linewidth]{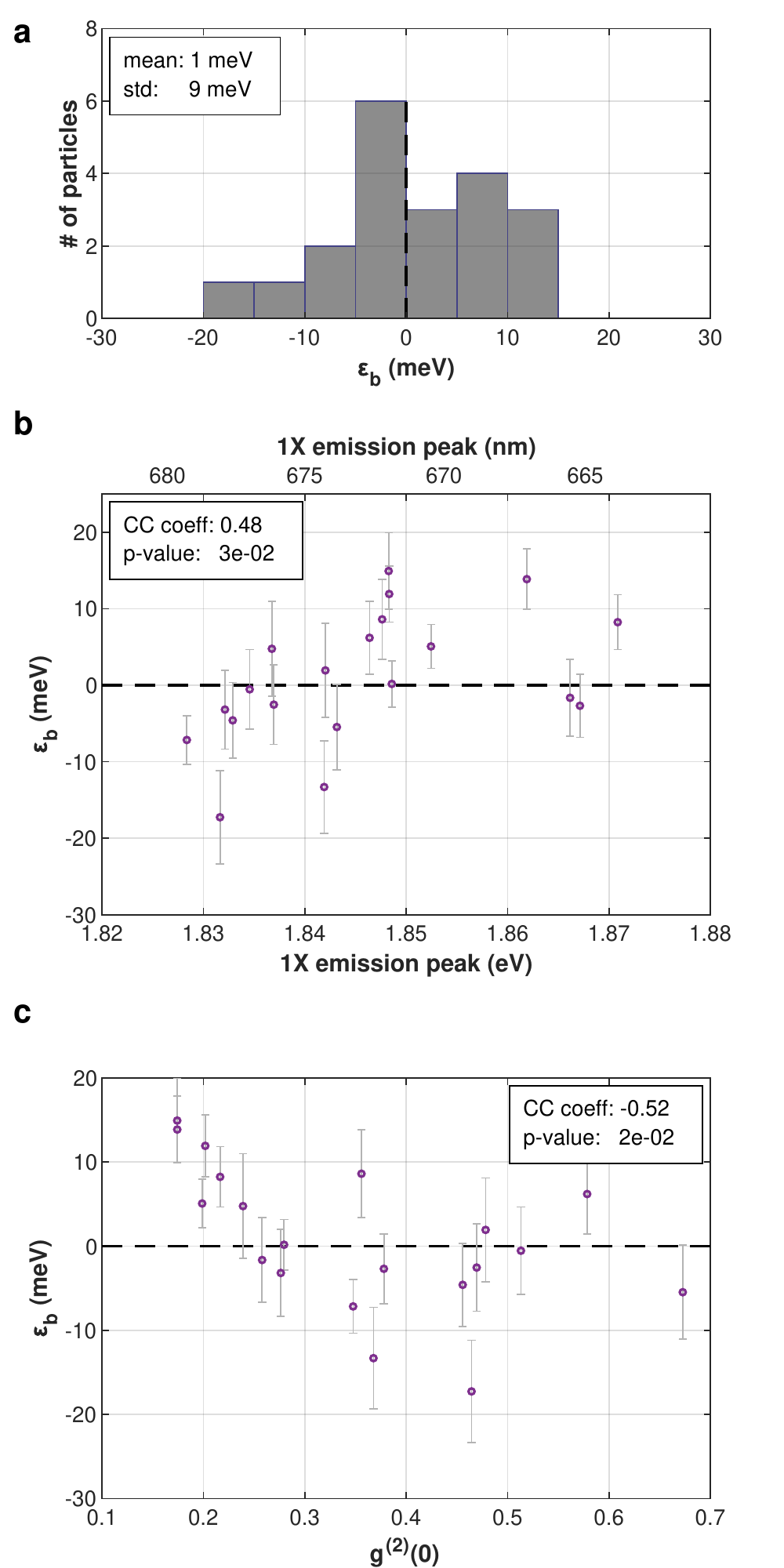}
    \caption{
    \textbf{\ce{CsPbI3} binding energy.} 
    \textbf{a)} Binding energy histogram for 20 NCs. Mean single-particle error is $\pm\SI{4.8}{meV}$. 
    \textbf{b)} Binding energy as a function of 1X emission peak. 
    \textbf{c)} Binding energy as a function of $g^{(2)}(0)$. 
    \textbf{std} - standard deviation, \textbf{CC coeff} - cross-correlation coefficient. \textbf{p-value} - p-value of the cross-correlation.
    }
    \label{fig:iod}
\end{figure}
% -----------------------------------------------------------------------------
% ------------------------------- main text -----------------------------------
% -----------------------------------------------------------------------------

\section*{Introduction}
Colloidal semiconductor nanocrystals (NC’s) have been extensively studied over the last three decades, owing to the ease of their synthesis and tunability of their photo-physical properties\cite{Efros2021}. Absorption of a photon by a NC leads to the formation of an exciton, a bound electron-hole pair, whose energy can be precisely tuned by varying the physical dimensions of the NC\cite{Brus1984}. In well passivated direct gap NCs, the dominant relaxation route of excitons is via photoluminescence (PL). Additional complexity is introduced when NCs are further excited, by absorbing multiple photons, to generate mutli-excitonic states. In the simplest of these states, the biexciton (BX), two excitons are confined within the NC.

PL from the BX state can serve as a probe to investigate exciton-exciton interaction within the NC. Relaxation from the BX to the singly-excited (1X) state can occur through radiative PL process or via non-radiative Auger processes\cite{Melnychuk2021}. Hence, the probability of radiative relaxation from the BX state, the BX quantum yield, is indicative of the relative rates of the two processes. A cascaded radiative relaxation from BX to 1X and further to the ground (G) state, results in the emission of two photons in rapid succession. The energy of the first photon ($E_{BX}$) will be shifted from the second ($E_{1X}$), according to exciton-exciton interaction. The value of this shift, the BX binding energy ($\varepsilon_b \equiv E_{1X} - E_{BX}$), is defined to be positive for attractive interaction. In intrinsic homogeneous or type-I NCs, where all charge-carriers are confined to the same volume, this interaction is typically attractive and stronger than in the bulk, due to the correlation energy of the confined excitons. In type-II heterostructure NCs, where electrons and holes are spatially separated, Coulombic repulsion of like-charged carriers can overwhelm this correlation energy, and result in an overall repulsive interaction\cite{Oron2007}. Significant effort has been directed at the evaluation and control of this value in II-VI and III-V semiconductor NCs, as it is critical to enhance their performance in various applications, such as sources of quantum light\cite{Senellart2017}, lasing media and LEDs\cite{Melnychuk2021} and photovoltaics\cite{Kramer2011}.

In recent years, there has been a surge of interest in lead halide perovskites (LHP) NCs of the form \ce{APbX3}, where A is a monovalent cation and X a halide anion. Their prominent features: near unity PL quantum yield, defect tolerance and tunable emission across the visible spectrum, have made them a promising candidate for various optoelectronic applications\cite{Protesescu2015,Kovalenko2017}. Additionally, at cryogenic temperatures, they exhibit long PL coherence times, which are desirable for emerging quantum optical technologies such as generation of coherent single-photons\cite{Utzat2019} and entangled photon pairs\cite{Akopian2006}. As in their II-VI and III-V counterparts, many of these applications stand to benefit, or even depend on reliable estimation of the BX binding energy.

However, the value of the BX binding energy in LHP NCs is a subject of current debate. Reported values for the prototypical all-inorganic \ce{CsPbBr3} NCs vary from +\SI{100}{meV}\cite{Castaneda2016} to \SI{-100}{meV}\cite{Dana2021}, while other experimental and theoretical works suggest a much lower bound of $\abs{\varepsilon_b}<\SI{20}{meV}$\cite{Shulenberger2019,Nguyen2020}. Common to all previous experimental works is their reliance on ensemble measurements. While these techniques proved invaluable in studying multiexcitonic states in NC, their ensemble nature introduces several possible sources for the estimation errors which may underlie the existing discrepancies. First, ensemble methods require fitting data to a model, and quantitative results often depend on the model chosen to analyze and interpret the data\cite{Makarov2016,Ashner2019}. In particular, the BX contribution might be hard to disentangle from other photo-physical or chemical processes such as charging or sintering\cite{Shulenberger2019,Lubin2021}, leading to ambiguities. Additionally, most methods require resolving the BX and 1X peaks spectrally, which might prove challenging when $\varepsilon_b$ is much smaller than the 1X homogeneous and inhomogeneous spectral broadening\cite{Shulenberger2019,Lubin2021}. Finally, the size heterogeneity, inherent to colloidally synthesized NC ensembles, can introduce systematic biases due to size dependent absorption cross-section of the 1X and BX states.

Room temperature single-particle heralded spectroscopy has been recently introduced as a way to overcome these shortcomings of ensemble approaches.\cite{Lubin2021,Vonk2021} This is achieved by temporally isolating photon pairs originating from the $BX{\rightarrow}1X{\rightarrow}G$ cascade of single-particles, and is hence free of all the aforementioned biases and ambiguities. In this letter, we utilize this technique to unambiguously determine the BX binding energies of the prototypical LHP NCs  \ce{CsPbBr3} and \ce{CsPbI3}. All \ce{CsPbBr3} single-particles measured featured an attractive exciton-exciton interaction ($\varepsilon_b=10\pm\SI{6}{meV}$), and a clear correlation of the BX binding energy with charge-carrier confinement was observed. Interestingly, \ce{CsPbI3} showed either weakly attractive or weakly repulsive exciton-exciton interaction with an average response around zero binding energy ($\varepsilon_b=1\pm\SI{9}{meV}$).

\section*{Results and Discussion}

\subsection*{Nanocrystals in this work}
Perovskite NCs investigated in this work were synthesised according to references\cite{Cao2020,Ahmed2018} (\ce{CsPbBr3}), and reference\cite{Pan2020} (\ce{CsPbI3}), with minor modifications (see \supp{1}). \ce{CsPbBr3} NCs, seen in \autoref{fig:particles}a, feature an edge size distribution of $5.9\pm\SI{1.3}{nm}$, \SI{2.44}{eV} ensemble emission peak and ${\sim}100\%$ quantum yield. For \ce{CsPbI3}, seen in \autoref{fig:particles}b, two size populations are visible. Smaller NCs (${\sim}80\%$ of the particles) with an edge size distribution of $7.2\pm\SI{1.9}{nm}$, and larger NCs with an edge size distribution of $15.4\pm\SI{3.3}{nm}$. The ensemble emission peak is at \SI{1.84}{eV} and the quantum yield is ${\sim}42\%$. Samples of isolated nanocrystals were prepared by spin coating a dilute solution of the NCs dispersed in a 3wt\% solution of poly(methylmetacrylate) (PMMA) in toluene on a glass coverslip.

\subsection*{Single-particle heralded spectroscopy.}
In order to measure the BX binding energy in single NCs, we use heralded spectroscopy - a technique that utilizes the temporal correlation of photon detections to unambiguously resolve the BX and 1X emission spectra. 
This technique was recently introduced and utilized to measure the BX binding energy of single CdSe/CdS/ZnS quantum dots at room temperature\cite{Lubin2021}. Briefly, an inverted microscope with a high numerical aperture objective is used to focus pulsed laser illumination on a single NC, and collect the emitted fluorescence. The collected fluorescence is then passed through a Czerny-Turner spectrometer with a single-photon avalanche diode (SPAD) array detector, so that each detected photon is time-stamped according to its arrival time, and energy-stamped according to the array pixel it was detected in (\autoref{fig:single}a). Post-selecting only photon pairs that follow the same excitation pulse, robustly isolates BX-1X emission cascades from emission of other overlapping emitting states, such as 1X or trions. The pump power is adjusted so that the average number of photons absorbed by a NC per pump pulse ($\ev{N}$) is low ($<0.4$, see \supp{2}). This helps prevent rapid deterioration of the NCs and minimize excitation of higher multiexcitonic states. A thorough description of the system and technique is given in reference\cite{Lubin2021}, and some modifications made to accommodate the different fluorescence parameters of the LHP NCs are described in \supp{3}.

\autoref{fig:single}b is a 2D-histogram of such post-selected photon pairs from a 5-minute measurement of a single \ce{CsPbBr3} NC. It shows the energy of the first arriving photon ($E_{BX}$) as a function of the second arriving photon ($E_{1X}$) of the pair. The green dashed line is a guide to the eye, marking same energy for both photons (undetectable by the system due to pixel dead time). The asymmetry of the histogram around this diagonal is indicative of an attractive exciton-exciton interaction ($E_{BX}$ is typically smaller than $E_{1X}$). This energy difference is quantified in \autoref{fig:single}c where the BX (red dots) and 1X (blue rings) spectra are extracted by full horizontal and full vertical binning, respectively, of \autoref{fig:single}b. This identification is corroborated by the good agreement between the 1X spectrum, and the spectrum of all detected photons (grey area, normalized).
The emission peaks of the BX and 1X spectra are estimated from fits to Cauchy-Lorentz distributions (matching color lines), and the BX binding energy is estimated as the difference in peak positions. For this specific NC, $\varepsilon_b = 13.5\pm\SI{1.8}{meV}$ (All errors in this paper are estimated as the 68\% confidence interval of the fit).

Two further insights were extracted from the same data-set. First, the normalized second order correlation of photon arrival times ($g^{(2)}(0)$), was calculated by the method described in reference\cite{Lubin2019}. This value is defined as the ratio between the number of detection pairs following the same excitation pulse and the expected number for a classical Poissonian emitter. The presence of the additional exciton in a doubly-excited NC increases the probability of nonradiative BX to 1X decay via Auger recombination. As a consequence, fewer photon pairs are emitted leading to $g^{(2)}(0)<1$, a phenomena termed photon anti-bunching. Hence, the value of $g^{(2)}(0)$ helps quantify the PL quantum yield of the BX state\cite{Nair2011}. 

Second, the \emph{time-gated} second order correlation of photon arrival times ($\Hat{g}^{(2)}(0)$) was calculated. This is performed by post-selecting only detections arriving within a time window of 1 to \SI{30}{ns} after any excitation pulse, and applying the same $g^{(2)}(0)$ calculation procedure to the resulting filtered data. Most multiexciton emission processes occur at timescales shorter than 1 ns (see \supp{4}), and are therefore filtered out by this time window. In single NCs, multi-exciton states are the only source for multiple photon detections following the same excitation pulse. Therefore, a low $\Hat{g}^{(2)}(0)$ is a good indication of whether the observed emission originates from a single particle or not\cite{Mangum2013,Benjamin2020}. ($g^{(2)}(0)$ and $\Hat{g}^{(2)}(0)$ are further discussed in \supp{5}). For this specific NC, $g^{(2)}(0) = 0.175\pm0.008$ and $\Hat{g}^{(2)}(0) = 0.012\pm0.003$.

\subsection*{\ce{CsPbBr3 NCs}.}
\autoref{fig:br}a shows the BX binding energy of 60 single \ce{CsPbBr3} NCs, determined using the procedure illustrated in \autoref{fig:single}. In our measurements, we maintain $\ev{N}\!\sim\!0.1$, and obtain a mean single-particle $\varepsilon_b$ error of $\pm\SI{3.1}{meV}$. To filter out accidental measurements of non-isolated NCs, only measurements where $\Hat{g}^{(2)}(0)<0.2$ are considered. All particles feature an attractive exciton-exciton interaction ($\varepsilon_b>0$), and the distribution is centered around $\varepsilon_b=10\pm\SI{6}{meV}$. \autoref{fig:br}b displays the binding energy of each NC as a function of the 1X emission peak position. A clear correlation between the two values is evident. This can be interpreted as the effect stronger charge-carrier confinement has on both the 1X emission peak (stronger confinement is correlated with higher energy emission peak) and the binding energy (stronger confinement is correlated with stronger interaction of the two excitons). The trend and magnitude are in reasonable agreement with theoretical predictions recently made by Nugyen \textit{et al.}\cite{Nguyen2020}, and bounds suggested by Shulenberger \textit{et al.}\cite{Shulenberger2019}.

The suggested interpretation is further corroborated in \autoref{fig:br}c. Here the BX binding energy is plotted as a function of $g^{(2)}(0)$, another value indicative of charge-carrier confinement. Namely, tighter confinement increases the rate of Auger processes\cite{Melnychuk2021}, and consequently reduces the yield of the competing radiative BX decay process, evident in lower $g^{(2)}(0)$. Therefore, the inverse correlation of $\varepsilon_b$ with $g^{(2)}(0)$ evident in \autoref{fig:br}c, can be seen as pointing to the same underlying correlation of the BX binding energy with charge-carrier confinement.

\subsection*{\ce{CsPbI3 NCs}.}
\ce{CsPbI3} NCs BX binding energies were measured by the same technique (\autoref{fig:iod}). Results feature $\varepsilon_b$ values distributed around zero ($\varepsilon_b=1\pm\SI{9}{meV}$, $\ev{N}\!\sim\!0.3$, mean single-particle error $\pm\SI{4.8}{meV}$). The trends observed for \ce{CsPbBr3} are visible here as well, where higher 1X emission peak energy and lower $g^{(2)}(0)$, or stronger confinement, are correlated with stronger attractive interaction (\autoref{fig:iod}b-c). As a result, while $\varepsilon_b$ values are mostly within reasonable error from zero, NCs featuring 1X emission peak lower (higher) than \SI{1.845}{eV} or $g^{(2)}(0)$ higher (lower) than 0.25 typically display a small negative (positive) $\varepsilon_b$ value. While the results are not as conclusive as for \ce{CsPbBr3}, they suggest that the weak exciton-exciton interaction in \ce{CsPbI3} NCs can be either repulsive or attractive, depending on the exact parameters of the single-particle.

It is noteworthy that \ce{CsPbI3} measurements were significantly more challenging than their \ce{CsPbBr3} counterparts. This is due to two factors. First, \ce{CsPbI3} NCs synthesized were typically less emissive and less stable under the conditions of our measurements. That resulted in many NCs deteriorating during the measurement (PL declines to near zero), before enough photon pairs were detected to extract reliable spectra fits. Second, current SPAD array technology is less sensitive at these longer wavelengths\cite{Antolovic2018}. The SPAD array detector used in this work has roughly twice the photon detection efficiency at the \ce{CsPbBr3} emission peak compared to the \ce{CsPbI3} emission peak. These two factors resulted in the smaller statistics and larger errors for \ce{CsPbI3} NCs in this work.

\subsection*{Discussion.}
The BX binding energies presented in this paper are at the lower range of values previously reported in the literature for these NCs (see a table of previously reported values in \supp{6}). While in some cases this might be attributed to the potential pitfalls associated with ensemble measurements discussed in the Introduction, it is also important to consider the possibility that heralded spectroscopy and ensemble measurements probe the NC in a qualitatively different excitation state.
For example, one widely adopted ensemble technique for estimating BX binding energy, involves recording the transient absorption (TA) spectrum of a probe pulse at very short ($<\SI{1}{ps}$) delays from a pump pulse i.e.\ before the relaxation of hot carriers to the band edge\cite{Klimov2007,Makarov2016,Aneesh2017}. The hot carrier pair generated by the pump shifts the spectral position of the absorption resonance for the probe photon, and this shift is recorded as the exciton-exciton interaction energy. In contrast, results presented in this paper rely on measurements of photon pairs emitted from individual $BX{\rightarrow}1X{\rightarrow}G$ cascades following the excitation pulse. Since the PL decay lifetimes of the BX and 1X states are significantly longer than the timescales of hot carrier relaxation in the NCs (see \supp{4}), our measurements probe the NC only after the hot carrier pairs have relaxed to the band edge.

Since the wavefunction of the hot exciton differs from that of a band-edge 1X state, the interaction energies may be different in the two cases. Studies on PbS nanocrystals indicate that the magnitude of interaction between a hot exciton and a band-edge exciton is larger than between two band-edge excitons \cite{Geiregat2014}. For \ce{CsPbI3} NCs, a recent study indicates that the estimated $\varepsilon_b$ increases as the pump wavelength decreases in short-delay TA experiments\cite{Yumoto2018}. Similar trends have been demonstrated for \ce{CsPbBr3} at cryogenic temperatures using two dimensional electron spectroscopy\cite{Huang2020}. In addition, analyses of TA measurements by Ashner \textit{et al.}, that do not employ short-delay spectra, did not result in large positive values (but rather in small negative values of few meV)\cite{Ashner2019}. Together, these observations suggest that $\varepsilon_b$ measured when both excitons are at the band edge would be lower than that measured when the first exciton is still hot. In this sense, heralded spectroscopy and short-delay TA are complementary measurements of band-edge and hot exciton-exciton interaction, respectively, and a careful comparison of the two can help uncover new insights into dynamics of exciton interactions in NCs. 

Negative $\varepsilon_b$ values, observed only for \ce{CsPbI3} NCs in this work, are less often reported in the literature for similar NCs (see \supp{6}). The origin of this repulsive interaction in not immediately apparent from existing theoretical models of intrinsic homogeneous semiconductor NCs\cite{Nguyen2020,Hu1990}. One possible explanation is a modification of the charge-carrier wavefunctions induced by surface ligands\cite{Baker2010}. This can result in a type-II potential landscape, where the electrons or the holes are localized at the NC surface, and Coulomb repulsion might dominate the exciton-exciton interaction. Alternatively, the electrostatic field generated by charge carriers trapped in the ligand-induced trap states can result in charge-separation, and a similar repulsive interaction. Another possibility, suggested by Ashner \textit{et al.}\cite{Ashner2019}, is the formation of polarons, supported by the deformable nature of the perovskite lattice. The results presented in this work cannot pinpoint a certain mechanism, and present limited statistics of negative $\varepsilon_b$ values. However, the apparent observation of a repulsive exciton-exciton interaction in a homogeneous nanocrystal highlights the importance of further investigating the effect of surface chemistry, environment and perovskite lattice on charge-carrier interaction in LHP NCs.

\section*{Conclusions}
Heralded spectroscopy enables us to unambiguously determine the biexciton binding energy ($\varepsilon_b$) of single lead halide perovskite nanocrystals. Using this technique, we demonstrate that \SI{{\sim}6}{nm} edge \ce{CsPbBr3} nanocrystals feature an attractive exciton-exciton interaction of $\varepsilon_b=10\pm\SI{6}{meV}$, which lies at the lower range of previously reported values. Interestingly, within the ensemble of \SI{{\sim}7}{nm} edge  \ce{CsPbI3} nanocrystals, some exhibit weak attractive interactions whereas in others weak exciton-exciton repulsion is observed. This rarely observed phenomenon in homogeneous nanocrystals warrants further investigation of charge-carrier interactions in these particles. In nanocrystals of both materials, the strength of attractive interaction exhibits a clear correlation with single-exciton emission peak position and photoluminescence anti-bunching ($g^{(2)}(0)$), highlighting the dependence of $\varepsilon_b$ on charge-carrier confinement. These insights into the physics of exciton interactions in lead halide perovskite nanocrystals can enable the developement of better engineered nanocrystals for future optoelectronic technologies. Moreover, the unprecedented ability to determine biexciton binding energy of single nanocrystals at room temperature is instrumental to their utilization in quantum technologies.

% -----------------------------------------------------------------------------
% ------------------------------- end matter ----------------------------------
% -----------------------------------------------------------------------------

\subsection*{Supporting Information}
Nanocrystal synthesis protocol; details of supporting analyses $\ev{N}$, $g^{(2)}(0)$ and $\Hat{g}^{(2)}(0)$; photoluminecence decay lifetime estimation; system parameters; list of previously reported biexciton binding energies.

\begin{acknowledgement}
The authors wish to thank Ron Tenne for his part in designing the SPAD array spectrometer. Financial support by the Crown center of Photonics and by the Minerva Foundation is gratefully acknowledged. DO is the incumbent of the Harry Weinrebe professorial chair of laser physics.
\end{acknowledgement}

\bibliography{main.bib}

\providecommand{\latin}[1]{#1}
\makeatletter
\providecommand{\doi}
  {\begingroup\let\do\@makeother\dospecials
  \catcode`\{=1 \catcode`\}=2 \doi@aux}
\providecommand{\doi@aux}[1]{\endgroup\texttt{#1}}
\makeatother
\providecommand*\mcitethebibliography{\thebibliography}
\csname @ifundefined\endcsname{endmcitethebibliography}
  {\let\endmcitethebibliography\endthebibliography}{}
\begin{mcitethebibliography}{34}
\providecommand*\natexlab[1]{#1}
\providecommand*\mciteSetBstSublistMode[1]{}
\providecommand*\mciteSetBstMaxWidthForm[2]{}
\providecommand*\mciteBstWouldAddEndPuncttrue
  {\def\EndOfBibitem{\unskip.}}
\providecommand*\mciteBstWouldAddEndPunctfalse
  {\let\EndOfBibitem\relax}
\providecommand*\mciteSetBstMidEndSepPunct[3]{}
\providecommand*\mciteSetBstSublistLabelBeginEnd[3]{}
\providecommand*\EndOfBibitem{}
\mciteSetBstSublistMode{f}
\mciteSetBstMaxWidthForm{subitem}{(\alph{mcitesubitemcount})}
\mciteSetBstSublistLabelBeginEnd
  {\mcitemaxwidthsubitemform\space}
  {\relax}
  {\relax}

\bibitem[Efros and Brus(2021)Efros, and Brus]{Efros2021}
Efros,~A.~L.; Brus,~L.~E. {Nanocrystal Quantum Dots: From Discovery to Modern
  Development}. \emph{ACS Nano} \textbf{2021}, \emph{15}, 6192--6210\relax
\mciteBstWouldAddEndPuncttrue
\mciteSetBstMidEndSepPunct{\mcitedefaultmidpunct}
{\mcitedefaultendpunct}{\mcitedefaultseppunct}\relax
\EndOfBibitem
\bibitem[Brus(1984)]{Brus1984}
Brus,~L.~E. {Electron-electron and electron-hole interactions in small
  semiconductor crystallites: The size dependence of the lowest excited
  electronic state}. \emph{The Journal of Chemical Physics} \textbf{1984},
  \emph{80}, 4403--4409\relax
\mciteBstWouldAddEndPuncttrue
\mciteSetBstMidEndSepPunct{\mcitedefaultmidpunct}
{\mcitedefaultendpunct}{\mcitedefaultseppunct}\relax
\EndOfBibitem
\bibitem[Melnychuk and Guyot-Sionnest(2021)Melnychuk, and
  Guyot-Sionnest]{Melnychuk2021}
Melnychuk,~C.; Guyot-Sionnest,~P. {Multicarrier dynamics in quantum dots}.
  \emph{Chemical Reviews} \textbf{2021}, \emph{121}, 2325--2372\relax
\mciteBstWouldAddEndPuncttrue
\mciteSetBstMidEndSepPunct{\mcitedefaultmidpunct}
{\mcitedefaultendpunct}{\mcitedefaultseppunct}\relax
\EndOfBibitem
\bibitem[Oron \latin{et~al.}(2007)Oron, Kazes, and Banin]{Oron2007}
Oron,~D.; Kazes,~M.; Banin,~U. {Multiexcitons in type-II colloidal
  semiconductor quantum dots}. \emph{Physical Review B - Condensed Matter and
  Materials Physics} \textbf{2007}, \emph{75}, 1--7\relax
\mciteBstWouldAddEndPuncttrue
\mciteSetBstMidEndSepPunct{\mcitedefaultmidpunct}
{\mcitedefaultendpunct}{\mcitedefaultseppunct}\relax
\EndOfBibitem
\bibitem[Senellart \latin{et~al.}(2017)Senellart, Solomon, and
  White]{Senellart2017}
Senellart,~P.; Solomon,~G.; White,~A. {High-performance semiconductor
  quantum-dot single-photon sources}. \emph{Nature Nanotechnology}
  \textbf{2017}, \emph{12}, 1026--1039\relax
\mciteBstWouldAddEndPuncttrue
\mciteSetBstMidEndSepPunct{\mcitedefaultmidpunct}
{\mcitedefaultendpunct}{\mcitedefaultseppunct}\relax
\EndOfBibitem
\bibitem[Kramer and Sargent(2011)Kramer, and Sargent]{Kramer2011}
Kramer,~I.~J.; Sargent,~E.~H. {Colloidal quantum dot photovoltaics: A path
  forward}. \emph{ACS Nano} \textbf{2011}, \emph{5}, 8506--8514\relax
\mciteBstWouldAddEndPuncttrue
\mciteSetBstMidEndSepPunct{\mcitedefaultmidpunct}
{\mcitedefaultendpunct}{\mcitedefaultseppunct}\relax
\EndOfBibitem
\bibitem[Protesescu \latin{et~al.}(2015)Protesescu, Yakunin, Bodnarchuk, Krieg,
  Caputo, Hendon, Yang, Walsh, and Kovalenko]{Protesescu2015}
Protesescu,~L.; Yakunin,~S.; Bodnarchuk,~M.~I.; Krieg,~F.; Caputo,~R.;
  Hendon,~C.~H.; Yang,~R.~X.; Walsh,~A.; Kovalenko,~M.~V. {Nanocrystals of
  Cesium Lead Halide Perovskites (CsPbX3, X = Cl, Br, and I): Novel
  Optoelectronic Materials Showing Bright Emission with Wide Color Gamut}.
  \emph{Nano Letters} \textbf{2015}, \emph{15}, 3692--3696\relax
\mciteBstWouldAddEndPuncttrue
\mciteSetBstMidEndSepPunct{\mcitedefaultmidpunct}
{\mcitedefaultendpunct}{\mcitedefaultseppunct}\relax
\EndOfBibitem
\bibitem[Kovalenko \latin{et~al.}(2017)Kovalenko, Protesescu, and
  Bodnarchuk]{Kovalenko2017}
Kovalenko,~M.~V.; Protesescu,~L.; Bodnarchuk,~M.~I. {Properties and potential
  optoelectronic applications of lead halide perovskite nanocrystals}.
  \emph{Science} \textbf{2017}, \emph{358}, 745--750\relax
\mciteBstWouldAddEndPuncttrue
\mciteSetBstMidEndSepPunct{\mcitedefaultmidpunct}
{\mcitedefaultendpunct}{\mcitedefaultseppunct}\relax
\EndOfBibitem
\bibitem[Utzat \latin{et~al.}(2019)Utzat, Sun, Kaplan, Krieg, Ginterseder,
  Spokoyny, Klein, Shulenberger, Perkinson, Kovalenko, and Bawendi]{Utzat2019}
Utzat,~H.; Sun,~W.; Kaplan,~A.~E.; Krieg,~F.; Ginterseder,~M.; Spokoyny,~B.;
  Klein,~N.~D.; Shulenberger,~K.~E.; Perkinson,~C.~F.; Kovalenko,~M.~V.;
  Bawendi,~M.~G. {Coherent single-photon emission from colloidal lead halide
  perovskite quantum dots}. \emph{Science} \textbf{2019}, \emph{363},
  1068--1072\relax
\mciteBstWouldAddEndPuncttrue
\mciteSetBstMidEndSepPunct{\mcitedefaultmidpunct}
{\mcitedefaultendpunct}{\mcitedefaultseppunct}\relax
\EndOfBibitem
\bibitem[Akopian \latin{et~al.}(2006)Akopian, Lindner, Poem, Berlatzky, Avron,
  Gershoni, Gerardot, and Petroff]{Akopian2006}
Akopian,~N.; Lindner,~N.~H.; Poem,~E.; Berlatzky,~Y.; Avron,~J.; Gershoni,~D.;
  Gerardot,~B.~D.; Petroff,~P.~M. {Entangled photon pairs from semiconductor
  quantum dots}. \emph{Physical Review Letters} \textbf{2006}, \emph{96},
  130501\relax
\mciteBstWouldAddEndPuncttrue
\mciteSetBstMidEndSepPunct{\mcitedefaultmidpunct}
{\mcitedefaultendpunct}{\mcitedefaultseppunct}\relax
\EndOfBibitem
\bibitem[Casta{\~{n}}eda \latin{et~al.}(2016)Casta{\~{n}}eda, Nagamine,
  Yassitepe, Bonato, Voznyy, Hoogland, Nogueira, Sargent, Cruz, and
  Padilha]{Castaneda2016}
Casta{\~{n}}eda,~J.~A.; Nagamine,~G.; Yassitepe,~E.; Bonato,~L.~G.; Voznyy,~O.;
  Hoogland,~S.; Nogueira,~A.~F.; Sargent,~E.~H.; Cruz,~C.~H.; Padilha,~L.~A.
  {Efficient Biexciton Interaction in Perovskite Quantum Dots under Weak and
  Strong Confinement}. \emph{ACS Nano} \textbf{2016}, \emph{10},
  8603--8609\relax
\mciteBstWouldAddEndPuncttrue
\mciteSetBstMidEndSepPunct{\mcitedefaultmidpunct}
{\mcitedefaultendpunct}{\mcitedefaultseppunct}\relax
\EndOfBibitem
\bibitem[Dana \latin{et~al.}(2021)Dana, Binyamin, Etgar, and Ruhman]{Dana2021}
Dana,~J.; Binyamin,~T.; Etgar,~L.; Ruhman,~S. {Unusually Strong Biexciton
  Repulsion Detected in Quantum Confined CsPbBr3Nanocrystals with Two and Three
  Pulse Femtosecond Spectroscopy}. \emph{ACS Nano} \textbf{2021}, \emph{15},
  23\relax
\mciteBstWouldAddEndPuncttrue
\mciteSetBstMidEndSepPunct{\mcitedefaultmidpunct}
{\mcitedefaultendpunct}{\mcitedefaultseppunct}\relax
\EndOfBibitem
\bibitem[Shulenberger \latin{et~al.}(2019)Shulenberger, Ashner, Ha, Krieg,
  Kovalenko, Tisdale, and Bawendi]{Shulenberger2019}
Shulenberger,~K.~E.; Ashner,~M.~N.; Ha,~S.~K.; Krieg,~F.; Kovalenko,~M.~V.;
  Tisdale,~W.~A.; Bawendi,~M.~G. {Setting an Upper Bound to the Biexciton
  Binding Energy in CsPbBr3 Perovskite Nanocrystals}. \emph{Journal of Physical
  Chemistry Letters} \textbf{2019}, \emph{10}, 5680--5686\relax
\mciteBstWouldAddEndPuncttrue
\mciteSetBstMidEndSepPunct{\mcitedefaultmidpunct}
{\mcitedefaultendpunct}{\mcitedefaultseppunct}\relax
\EndOfBibitem
\bibitem[Nguyen \latin{et~al.}(2020)Nguyen, Blundell, and Guet]{Nguyen2020}
Nguyen,~T.~P.; Blundell,~S.~A.; Guet,~C. {Calculation of the biexciton shift in
  nanocrystals of inorganic perovskites}. \emph{Physical Review B}
  \textbf{2020}, \emph{101}, 125424\relax
\mciteBstWouldAddEndPuncttrue
\mciteSetBstMidEndSepPunct{\mcitedefaultmidpunct}
{\mcitedefaultendpunct}{\mcitedefaultseppunct}\relax
\EndOfBibitem
\bibitem[Makarov \latin{et~al.}(2016)Makarov, Guo, Isaienko, Liu, Robel, and
  Klimov]{Makarov2016}
Makarov,~N.~S.; Guo,~S.; Isaienko,~O.; Liu,~W.; Robel,~I.; Klimov,~V.~I.
  {Spectral and Dynamical Properties of Single Excitons, Biexcitons, and Trions
  in Cesium-Lead-Halide Perovskite Quantum Dots}. \emph{Nano Letters}
  \textbf{2016}, \emph{16}, 2349--2362\relax
\mciteBstWouldAddEndPuncttrue
\mciteSetBstMidEndSepPunct{\mcitedefaultmidpunct}
{\mcitedefaultendpunct}{\mcitedefaultseppunct}\relax
\EndOfBibitem
\bibitem[Ashner \latin{et~al.}(2019)Ashner, Shulenberger, Krieg, Powers,
  Kovalenko, Bawendi, and Tisdale]{Ashner2019}
Ashner,~M.~N.; Shulenberger,~K.~E.; Krieg,~F.; Powers,~E.~R.; Kovalenko,~M.~V.;
  Bawendi,~M.~G.; Tisdale,~W.~A. {Size-dependent biexciton spectrum in cspbbr3
  perovskite nanocrystals}. \emph{ACS Energy Letters} \textbf{2019}, \emph{4},
  2639--2645\relax
\mciteBstWouldAddEndPuncttrue
\mciteSetBstMidEndSepPunct{\mcitedefaultmidpunct}
{\mcitedefaultendpunct}{\mcitedefaultseppunct}\relax
\EndOfBibitem
\bibitem[Lubin \latin{et~al.}(2021)Lubin, Tenne, Ulku, Antolovic, Burri, Karg,
  Yallapragada, Bruschini, Charbon, and Oron]{Lubin2021}
Lubin,~G.; Tenne,~R.; Ulku,~A.~C.; Antolovic,~I.~M.; Burri,~S.; Karg,~S.;
  Yallapragada,~V.~J.; Bruschini,~C.; Charbon,~E.; Oron,~D. {Heralded
  spectroscopy reveals exciton-exciton correlations in single colloidal quantum
  dots}. \emph{arXiv} \textbf{2021}, \relax
\mciteBstWouldAddEndPunctfalse
\mciteSetBstMidEndSepPunct{\mcitedefaultmidpunct}
{}{\mcitedefaultseppunct}\relax
\EndOfBibitem
\bibitem[Vonk \latin{et~al.}(2021)Vonk, Heemskerk, Keitel, Hinterding,
  Geuchies, Houtepen, and Rabouw]{Vonk2021}
Vonk,~S. J.~W.; Heemskerk,~B. A.~J.; Keitel,~R.~C.; Hinterding,~S. O.~M.;
  Geuchies,~J.~J.; Houtepen,~A.~J.; Rabouw,~F.~T. {Biexciton Binding Energy and
  Line width of Single Quantum Dots at Room Temperature}. \emph{Nano Letters}
  \textbf{2021}, \relax
\mciteBstWouldAddEndPunctfalse
\mciteSetBstMidEndSepPunct{\mcitedefaultmidpunct}
{}{\mcitedefaultseppunct}\relax
\EndOfBibitem
\bibitem[Cao \latin{et~al.}(2020)Cao, Zhu, Li, Zhang, Chen, Lin, and
  Zhu]{Cao2020}
Cao,~Y.; Zhu,~W.; Li,~L.; Zhang,~Z.; Chen,~Z.; Lin,~Y.; Zhu,~J.~J.
  {Size-selected and surface-passivated CsPbBr3 perovskite nanocrystals for
  self-enhanced electrochemiluminescence in aqueous media}. \emph{Nanoscale}
  \textbf{2020}, \emph{12}, 7321--7329\relax
\mciteBstWouldAddEndPuncttrue
\mciteSetBstMidEndSepPunct{\mcitedefaultmidpunct}
{\mcitedefaultendpunct}{\mcitedefaultseppunct}\relax
\EndOfBibitem
\bibitem[Ahmed \latin{et~al.}(2018)Ahmed, Seth, and Samanta]{Ahmed2018}
Ahmed,~T.; Seth,~S.; Samanta,~A. {Boosting the Photoluminescence of CsPbX3 (X =
  Cl, Br, I) Perovskite Nanocrystals Covering a Wide Wavelength Range by
  Postsynthetic Treatment with Tetrafluoroborate Salts}. \emph{Chemistry of
  Materials} \textbf{2018}, \emph{30}, 3633--3637\relax
\mciteBstWouldAddEndPuncttrue
\mciteSetBstMidEndSepPunct{\mcitedefaultmidpunct}
{\mcitedefaultendpunct}{\mcitedefaultseppunct}\relax
\EndOfBibitem
\bibitem[Pan \latin{et~al.}(2020)Pan, Zhang, Qi, Conkle, Han, Zhu, Box,
  Shahbazyan, and Dai]{Pan2020}
Pan,~L.; Zhang,~L.; Qi,~Y.; Conkle,~K.; Han,~F.; Zhu,~X.; Box,~D.;
  Shahbazyan,~T.~V.; Dai,~Q. {Stable CsPbI3Nanocrystals Modified by Tetra-
  n-butylammonium Iodide for Light-Emitting Diodes}. \emph{ACS Applied Nano
  Materials} \textbf{2020}, \emph{3}, 9260--9267\relax
\mciteBstWouldAddEndPuncttrue
\mciteSetBstMidEndSepPunct{\mcitedefaultmidpunct}
{\mcitedefaultendpunct}{\mcitedefaultseppunct}\relax
\EndOfBibitem
\bibitem[Lubin \latin{et~al.}(2019)Lubin, Tenne, Antolovic, Charbon, Bruschini,
  and Oron]{Lubin2019}
Lubin,~G.; Tenne,~R.; Antolovic,~I.~M.; Charbon,~E.; Bruschini,~C.; Oron,~D.
  {Quantum correlation measurement with single photon avalanche diode arrays}.
  \textbf{2019}, \emph{27}, 32863--32882\relax
\mciteBstWouldAddEndPuncttrue
\mciteSetBstMidEndSepPunct{\mcitedefaultmidpunct}
{\mcitedefaultendpunct}{\mcitedefaultseppunct}\relax
\EndOfBibitem
\bibitem[Nair \latin{et~al.}(2011)Nair, Zhao, and Bawendi]{Nair2011}
Nair,~G.; Zhao,~J.; Bawendi,~M.~G. {Biexciton quantum yield of single
  semiconductor nanocrystals from photon statistics}. \emph{Nano Letters}
  \textbf{2011}, \emph{11}, 1136--1140\relax
\mciteBstWouldAddEndPuncttrue
\mciteSetBstMidEndSepPunct{\mcitedefaultmidpunct}
{\mcitedefaultendpunct}{\mcitedefaultseppunct}\relax
\EndOfBibitem
\bibitem[Mangum \latin{et~al.}(2013)Mangum, Ghosh, Hollingsworth, and
  Htoon]{Mangum2013}
Mangum,~B.~D.; Ghosh,~Y.; Hollingsworth,~J.~A.; Htoon,~H. {Disentangling the
  effects of clustering and multi-exciton emission in second-order photon
  correlation experiments}. \emph{Optics Express} \textbf{2013}, \emph{21},
  7419\relax
\mciteBstWouldAddEndPuncttrue
\mciteSetBstMidEndSepPunct{\mcitedefaultmidpunct}
{\mcitedefaultendpunct}{\mcitedefaultseppunct}\relax
\EndOfBibitem
\bibitem[Benjamin \latin{et~al.}(2020)Benjamin, Yallapragada, Amgar, Yang,
  Tenne, and Oron]{Benjamin2020}
Benjamin,~E.; Yallapragada,~V.~J.; Amgar,~D.; Yang,~G.; Tenne,~R.; Oron,~D.
  {Temperature Dependence of Excitonic and Biexcitonic Decay Rates in Colloidal
  Nanoplatelets by Time-Gated Photon Correlation}. \emph{Journal of Physical
  Chemistry Letters} \textbf{2020}, \emph{11}, 6513--6518\relax
\mciteBstWouldAddEndPuncttrue
\mciteSetBstMidEndSepPunct{\mcitedefaultmidpunct}
{\mcitedefaultendpunct}{\mcitedefaultseppunct}\relax
\EndOfBibitem
\bibitem[Antolovic \latin{et~al.}(2018)Antolovic, Bruschini, and
  Charbon]{Antolovic2018}
Antolovic,~I.~M.; Bruschini,~C.; Charbon,~E. {Dynamic range extension for
  photon counting arrays}. \emph{Optics Express} \textbf{2018}, \emph{26},
  22234\relax
\mciteBstWouldAddEndPuncttrue
\mciteSetBstMidEndSepPunct{\mcitedefaultmidpunct}
{\mcitedefaultendpunct}{\mcitedefaultseppunct}\relax
\EndOfBibitem
\bibitem[Klimov(2007)]{Klimov2007}
Klimov,~V.~I. {Spectral and dynamical properties of multiexcitons in
  semiconductor nanocrystals}. \emph{Annual Review of Physical Chemistry}
  \textbf{2007}, \emph{58}, 635--673\relax
\mciteBstWouldAddEndPuncttrue
\mciteSetBstMidEndSepPunct{\mcitedefaultmidpunct}
{\mcitedefaultendpunct}{\mcitedefaultseppunct}\relax
\EndOfBibitem
\bibitem[Aneesh \latin{et~al.}(2017)Aneesh, Swarnkar, Kumar~Ravi, Sharma, Nag,
  and Adarsh]{Aneesh2017}
Aneesh,~J.; Swarnkar,~A.; Kumar~Ravi,~V.; Sharma,~R.; Nag,~A.; Adarsh,~K.~V.
  {Ultrafast Exciton Dynamics in Colloidal CsPbBr3 Perovskite Nanocrystals:
  Biexciton Effect and Auger Recombination}. \emph{Journal of Physical
  Chemistry C} \textbf{2017}, \emph{121}, 4734--4739\relax
\mciteBstWouldAddEndPuncttrue
\mciteSetBstMidEndSepPunct{\mcitedefaultmidpunct}
{\mcitedefaultendpunct}{\mcitedefaultseppunct}\relax
\EndOfBibitem
\bibitem[Geiregat \latin{et~al.}(2014)Geiregat, Houtepen, Justo, Grozema,
  Van~Thourhout, and Hens]{Geiregat2014}
Geiregat,~P.; Houtepen,~A.; Justo,~Y.; Grozema,~F.~C.; Van~Thourhout,~D.;
  Hens,~Z. {Coulomb shifts upon exciton addition to photoexcited PbS colloidal
  quantum Dots}. \emph{Journal of Physical Chemistry C} \textbf{2014},
  \emph{118}, 22284--22290\relax
\mciteBstWouldAddEndPuncttrue
\mciteSetBstMidEndSepPunct{\mcitedefaultmidpunct}
{\mcitedefaultendpunct}{\mcitedefaultseppunct}\relax
\EndOfBibitem
\bibitem[Yumoto \latin{et~al.}(2018)Yumoto, Tahara, Kawawaki, Saruyama, Sato,
  Teranishi, and Kanemitsu]{Yumoto2018}
Yumoto,~G.; Tahara,~H.; Kawawaki,~T.; Saruyama,~M.; Sato,~R.; Teranishi,~T.;
  Kanemitsu,~Y. {Hot Biexciton Effect on Optical Gain in CsPbI3 Perovskite
  Nanocrystals}. \emph{Journal of Physical Chemistry Letters} \textbf{2018},
  \emph{9}, 2222--2228\relax
\mciteBstWouldAddEndPuncttrue
\mciteSetBstMidEndSepPunct{\mcitedefaultmidpunct}
{\mcitedefaultendpunct}{\mcitedefaultseppunct}\relax
\EndOfBibitem
\bibitem[Huang \latin{et~al.}(2020)Huang, Chen, Zhang, Qin, Yu, Wang, and
  Xiao]{Huang2020}
Huang,~X.; Chen,~L.; Zhang,~C.; Qin,~Z.; Yu,~B.; Wang,~X.; Xiao,~M.
  {Inhomogeneous Biexciton Binding in Perovskite Semiconductor Nanocrystals
  Measured with Two-Dimensional Spectroscopy}. \emph{Journal of Physical
  Chemistry Letters} \textbf{2020}, \emph{11}, 10173--10181\relax
\mciteBstWouldAddEndPuncttrue
\mciteSetBstMidEndSepPunct{\mcitedefaultmidpunct}
{\mcitedefaultendpunct}{\mcitedefaultseppunct}\relax
\EndOfBibitem
\bibitem[Hu \latin{et~al.}(1990)Hu, Koch, Lindberg, Peyghambarian, Pollock, and
  Abraham]{Hu1990}
Hu,~Y.~Z.; Koch,~S.~W.; Lindberg,~M.; Peyghambarian,~N.; Pollock,~E.~L.;
  Abraham,~F.~F. {Biexcitons in semiconductor quantum dots}. \emph{Physical
  Review Letters} \textbf{1990}, \emph{64}, 1805--1807\relax
\mciteBstWouldAddEndPuncttrue
\mciteSetBstMidEndSepPunct{\mcitedefaultmidpunct}
{\mcitedefaultendpunct}{\mcitedefaultseppunct}\relax
\EndOfBibitem
\bibitem[Baker and Kamat(2010)Baker, and Kamat]{Baker2010}
Baker,~D.~R.; Kamat,~P.~V. {Tuning the emission of CdSe quantum dots by
  controlled trap enhancement}. \emph{Langmuir} \textbf{2010}, \emph{26},
  11272--11276\relax
\mciteBstWouldAddEndPuncttrue
\mciteSetBstMidEndSepPunct{\mcitedefaultmidpunct}
{\mcitedefaultendpunct}{\mcitedefaultseppunct}\relax
\EndOfBibitem
\end{mcitethebibliography}


\providecommand{\latin}[1]{#1}
\makeatletter
\providecommand{\doi}
  {\begingroup\let\do\@makeother\dospecials
  \catcode`\{=1 \catcode`\}=2 \doi@aux}
\providecommand{\doi@aux}[1]{\endgroup\texttt{#1}}
\makeatother
\providecommand*\mcitethebibliography{\thebibliography}
\csname @ifundefined\endcsname{endmcitethebibliography}
  {\let\endmcitethebibliography\endthebibliography}{}
\begin{mcitethebibliography}{16}
\providecommand*\natexlab[1]{#1}
\providecommand*\mciteSetBstSublistMode[1]{}
\providecommand*\mciteSetBstMaxWidthForm[2]{}
\providecommand*\mciteBstWouldAddEndPuncttrue
  {\def\EndOfBibitem{\unskip.}}
\providecommand*\mciteBstWouldAddEndPunctfalse
  {\let\EndOfBibitem\relax}
\providecommand*\mciteSetBstMidEndSepPunct[3]{}
\providecommand*\mciteSetBstSublistLabelBeginEnd[3]{}
\providecommand*\EndOfBibitem{}
\mciteSetBstSublistMode{f}
\mciteSetBstMaxWidthForm{subitem}{(\alph{mcitesubitemcount})}
\mciteSetBstSublistLabelBeginEnd
  {\mcitemaxwidthsubitemform\space}
  {\relax}
  {\relax}

\bibitem[Cao \latin{et~al.}(2020)Cao, Zhu, Li, Zhang, Chen, Lin, and
  Zhu]{Cao2020}
Cao,~Y.; Zhu,~W.; Li,~L.; Zhang,~Z.; Chen,~Z.; Lin,~Y.; Zhu,~J.~J.
  {Size-selected and surface-passivated CsPbBr3 perovskite nanocrystals for
  self-enhanced electrochemiluminescence in aqueous media}. \emph{Nanoscale}
  \textbf{2020}, \emph{12}, 7321--7329\relax
\mciteBstWouldAddEndPuncttrue
\mciteSetBstMidEndSepPunct{\mcitedefaultmidpunct}
{\mcitedefaultendpunct}{\mcitedefaultseppunct}\relax
\EndOfBibitem
\bibitem[Ahmed \latin{et~al.}(2018)Ahmed, Seth, and Samanta]{Ahmed2018}
Ahmed,~T.; Seth,~S.; Samanta,~A. {Boosting the Photoluminescence of CsPbX3 (X =
  Cl, Br, I) Perovskite Nanocrystals Covering a Wide Wavelength Range by
  Postsynthetic Treatment with Tetrafluoroborate Salts}. \emph{Chemistry of
  Materials} \textbf{2018}, \emph{30}, 3633--3637\relax
\mciteBstWouldAddEndPuncttrue
\mciteSetBstMidEndSepPunct{\mcitedefaultmidpunct}
{\mcitedefaultendpunct}{\mcitedefaultseppunct}\relax
\EndOfBibitem
\bibitem[Pan \latin{et~al.}(2020)Pan, Zhang, Qi, Conkle, Han, Zhu, Box,
  Shahbazyan, and Dai]{Pan2020}
Pan,~L.; Zhang,~L.; Qi,~Y.; Conkle,~K.; Han,~F.; Zhu,~X.; Box,~D.;
  Shahbazyan,~T.~V.; Dai,~Q. {Stable CsPbI3Nanocrystals Modified by Tetra-
  n-butylammonium Iodide for Light-Emitting Diodes}. \emph{ACS Applied Nano
  Materials} \textbf{2020}, \emph{3}, 9260--9267\relax
\mciteBstWouldAddEndPuncttrue
\mciteSetBstMidEndSepPunct{\mcitedefaultmidpunct}
{\mcitedefaultendpunct}{\mcitedefaultseppunct}\relax
\EndOfBibitem
\bibitem[Lubin \latin{et~al.}(2021)Lubin, Tenne, Ulku, Antolovic, Burri, Karg,
  Yallapragada, Bruschini, Charbon, and Oron]{Lubin2021}
Lubin,~G.; Tenne,~R.; Ulku,~A.~C.; Antolovic,~I.~M.; Burri,~S.; Karg,~S.;
  Yallapragada,~V.~J.; Bruschini,~C.; Charbon,~E.; Oron,~D. {Heralded
  spectroscopy reveals exciton-exciton correlations in single colloidal quantum
  dots}. \emph{arXiv} \textbf{2021}, \relax
\mciteBstWouldAddEndPunctfalse
\mciteSetBstMidEndSepPunct{\mcitedefaultmidpunct}
{}{\mcitedefaultseppunct}\relax
\EndOfBibitem
\bibitem[De~Jong \latin{et~al.}(2017)De~Jong, Yamashita, Gomez, Ashida,
  Fujiwara, and Gregorkiewicz]{DeJong2017}
De~Jong,~E.~M.; Yamashita,~G.; Gomez,~L.; Ashida,~M.; Fujiwara,~Y.;
  Gregorkiewicz,~T. {Multiexciton lifetime in all-inorganic CsPbBr3 perovskite
  nanocrystals}. \emph{Journal of Physical Chemistry C} \textbf{2017},
  \emph{121}, 1941--1947\relax
\mciteBstWouldAddEndPuncttrue
\mciteSetBstMidEndSepPunct{\mcitedefaultmidpunct}
{\mcitedefaultendpunct}{\mcitedefaultseppunct}\relax
\EndOfBibitem
\bibitem[Lubin \latin{et~al.}(2019)Lubin, Tenne, Antolovic, Charbon, Bruschini,
  and Oron]{Lubin2019}
Lubin,~G.; Tenne,~R.; Antolovic,~I.~M.; Charbon,~E.; Bruschini,~C.; Oron,~D.
  {Quantum correlation measurement with single photon avalanche diode arrays}.
  \textbf{2019}, \emph{27}, 32863--32882\relax
\mciteBstWouldAddEndPuncttrue
\mciteSetBstMidEndSepPunct{\mcitedefaultmidpunct}
{\mcitedefaultendpunct}{\mcitedefaultseppunct}\relax
\EndOfBibitem
\bibitem[Casta{\~{n}}eda \latin{et~al.}(2016)Casta{\~{n}}eda, Nagamine,
  Yassitepe, Bonato, Voznyy, Hoogland, Nogueira, Sargent, Cruz, and
  Padilha]{Castaneda2016}
Casta{\~{n}}eda,~J.~A.; Nagamine,~G.; Yassitepe,~E.; Bonato,~L.~G.; Voznyy,~O.;
  Hoogland,~S.; Nogueira,~A.~F.; Sargent,~E.~H.; Cruz,~C.~H.; Padilha,~L.~A.
  {Efficient Biexciton Interaction in Perovskite Quantum Dots under Weak and
  Strong Confinement}. \emph{ACS Nano} \textbf{2016}, \emph{10},
  8603--8609\relax
\mciteBstWouldAddEndPuncttrue
\mciteSetBstMidEndSepPunct{\mcitedefaultmidpunct}
{\mcitedefaultendpunct}{\mcitedefaultseppunct}\relax
\EndOfBibitem
\bibitem[Ashner \latin{et~al.}(2019)Ashner, Shulenberger, Krieg, Powers,
  Kovalenko, Bawendi, and Tisdale]{Ashner2019}
Ashner,~M.~N.; Shulenberger,~K.~E.; Krieg,~F.; Powers,~E.~R.; Kovalenko,~M.~V.;
  Bawendi,~M.~G.; Tisdale,~W.~A. {Size-dependent biexciton spectrum in cspbbr3
  perovskite nanocrystals}. \emph{ACS Energy Letters} \textbf{2019}, \emph{4},
  2639--2645\relax
\mciteBstWouldAddEndPuncttrue
\mciteSetBstMidEndSepPunct{\mcitedefaultmidpunct}
{\mcitedefaultendpunct}{\mcitedefaultseppunct}\relax
\EndOfBibitem
\bibitem[Makarov \latin{et~al.}(2016)Makarov, Guo, Isaienko, Liu, Robel, and
  Klimov]{Makarov2016}
Makarov,~N.~S.; Guo,~S.; Isaienko,~O.; Liu,~W.; Robel,~I.; Klimov,~V.~I.
  {Spectral and Dynamical Properties of Single Excitons, Biexcitons, and Trions
  in Cesium-Lead-Halide Perovskite Quantum Dots}. \emph{Nano Letters}
  \textbf{2016}, \emph{16}, 2349--2362\relax
\mciteBstWouldAddEndPuncttrue
\mciteSetBstMidEndSepPunct{\mcitedefaultmidpunct}
{\mcitedefaultendpunct}{\mcitedefaultseppunct}\relax
\EndOfBibitem
\bibitem[Aneesh \latin{et~al.}(2017)Aneesh, Swarnkar, Kumar~Ravi, Sharma, Nag,
  and Adarsh]{Aneesh2017}
Aneesh,~J.; Swarnkar,~A.; Kumar~Ravi,~V.; Sharma,~R.; Nag,~A.; Adarsh,~K.~V.
  {Ultrafast Exciton Dynamics in Colloidal CsPbBr3 Perovskite Nanocrystals:
  Biexciton Effect and Auger Recombination}. \emph{Journal of Physical
  Chemistry C} \textbf{2017}, \emph{121}, 4734--4739\relax
\mciteBstWouldAddEndPuncttrue
\mciteSetBstMidEndSepPunct{\mcitedefaultmidpunct}
{\mcitedefaultendpunct}{\mcitedefaultseppunct}\relax
\EndOfBibitem
\bibitem[Yin \latin{et~al.}(2017)Yin, Chen, Song, Lv, Hu, Sun, Yu, Zhang, Wang,
  Zhang, and Xiao]{Yin2017}
Yin,~C.; Chen,~L.; Song,~N.; Lv,~Y.; Hu,~F.; Sun,~C.; Yu,~W.~W.; Zhang,~C.;
  Wang,~X.; Zhang,~Y.; Xiao,~M. {Bright-Exciton Fine-Structure Splittings in
  Single Perovskite Nanocrystals}. \emph{Physical Review Letters}
  \textbf{2017}, \emph{119}, 026401\relax
\mciteBstWouldAddEndPuncttrue
\mciteSetBstMidEndSepPunct{\mcitedefaultmidpunct}
{\mcitedefaultendpunct}{\mcitedefaultseppunct}\relax
\EndOfBibitem
\bibitem[Yumoto \latin{et~al.}(2018)Yumoto, Tahara, Kawawaki, Saruyama, Sato,
  Teranishi, and Kanemitsu]{Yumoto2018}
Yumoto,~G.; Tahara,~H.; Kawawaki,~T.; Saruyama,~M.; Sato,~R.; Teranishi,~T.;
  Kanemitsu,~Y. {Hot Biexciton Effect on Optical Gain in CsPbI3 Perovskite
  Nanocrystals}. \emph{Journal of Physical Chemistry Letters} \textbf{2018},
  \emph{9}, 2222--2228\relax
\mciteBstWouldAddEndPuncttrue
\mciteSetBstMidEndSepPunct{\mcitedefaultmidpunct}
{\mcitedefaultendpunct}{\mcitedefaultseppunct}\relax
\EndOfBibitem
\bibitem[Huang \latin{et~al.}(2020)Huang, Chen, Zhang, Qin, Yu, Wang, and
  Xiao]{Huang2020}
Huang,~X.; Chen,~L.; Zhang,~C.; Qin,~Z.; Yu,~B.; Wang,~X.; Xiao,~M.
  {Inhomogeneous Biexciton Binding in Perovskite Semiconductor Nanocrystals
  Measured with Two-Dimensional Spectroscopy}. \emph{Journal of Physical
  Chemistry Letters} \textbf{2020}, \emph{11}, 10173--10181\relax
\mciteBstWouldAddEndPuncttrue
\mciteSetBstMidEndSepPunct{\mcitedefaultmidpunct}
{\mcitedefaultendpunct}{\mcitedefaultseppunct}\relax
\EndOfBibitem
\bibitem[An \latin{et~al.}(2021)An, Pan, Shen, Wang, Geng, Li, Zhao, Sun, and
  Wu]{Shen2021}
An,~L.; Pan,~K.; Shen,~X.; Wang,~S.; Geng,~C.; Li,~L.; Zhao,~E.; Sun,~J.;
  Wu,~W. {Red shift of bleaching signals in femtosecond transient absorption
  spectra of CsPbX3(X = Cl/Br, Br, Br/I) nanocrystals induced by the biexciton
  effect}. \emph{Journal of Physical Chemistry C} \textbf{2021}, \emph{125},
  5278--5287\relax
\mciteBstWouldAddEndPuncttrue
\mciteSetBstMidEndSepPunct{\mcitedefaultmidpunct}
{\mcitedefaultendpunct}{\mcitedefaultseppunct}\relax
\EndOfBibitem
\bibitem[Dana \latin{et~al.}(2021)Dana, Binyamin, Etgar, and Ruhman]{Dana2021}
Dana,~J.; Binyamin,~T.; Etgar,~L.; Ruhman,~S. {Unusually Strong Biexciton
  Repulsion Detected in Quantum Confined CsPbBr3Nanocrystals with Two and Three
  Pulse Femtosecond Spectroscopy}. \emph{ACS Nano} \textbf{2021}, \emph{15},
  23\relax
\mciteBstWouldAddEndPuncttrue
\mciteSetBstMidEndSepPunct{\mcitedefaultmidpunct}
{\mcitedefaultendpunct}{\mcitedefaultseppunct}\relax
\EndOfBibitem
\end{mcitethebibliography}

\end{document}

\twocolumn[
\begin{@twocolumnfalse}
\oldmaketitle
\begin{abstract}
This supporting information describes in greater detail the synthesis, data analysis and system parameters, as well as provides some additional information to the work described in ``Resolving the controversy in biexciton binding energy of cesium lead halide perovskite nanocrystals through heralded single-particle spectroscopy". Sections are brought in the order of their reference in the main text: nanocrystal synthesis protocol; details of supporting analyses $\ev{N}$, $g^{(2)}(0)$ and $\Hat{g^{(2)}(0)}$; photoluminecence decay lifetime estimation; system parameters; list of previously reported biexciton binding energies.     
\end{abstract}
\end{@twocolumnfalse}
]

% -----------------------------------------------------------------------------
% ------------------------------- main text -----------------------------------
% -----------------------------------------------------------------------------

\section{Synthesis protocol}
This section describes the synthesis protocol of the \ce{CsPbBr3} and \ce{CsPbI3} nanocrystals (NC) used in this work.

\subsection{Materials}
\ce{Cs2CO3} (99.995\%, Sigma-Aldrich), octadecene (ODE, 90\%, Sigma-Aldrich), oleic acid (OA, 90\%, Sigma-Aldrich), oleylamine (OLA, 70\%, Sigma-Aldrich), \ce{PbBr2} (98\%, Sigma-Aldrich), \ce{PbI2} (99\%, Aldrich), toluene (99.8\%, Sigma-Aldrich, anhydrous), hexane (99.5\%, Sigma-Aldrich, anhydrous) ammonium tetrafluoroborate (\ce{NH4BF4}, 99.999\%, Sigma-Aldrich), tetradecylphosphonic acid (TDPA, 99\%, Sigma-Aldrich), Trioctylphosphine oxide (TOPO, 90\%, Sigma-Aldrich)

\subsection{Cs-Oleate preparation}
\ce{Cs2CO3} (\SI{101.7}{mg}), OA (\SI{312.5}{\micro\liter}) and ODE (\SI{5}{mL}) were mixed in a \SI{50}{mL} round bottom flask, heated at \SI{120}{\celsius} under vacuum for one hour. Then the temperature was raised to \SI{160}{\celsius} and the mixture was kept for \SI{10}{\minute} under Ar atmosphere. For the injection procedure, Cs-oleate was kept at \SI{120}{\celsius} under Ar.

\subsection{Synthesis of \ce{CsPbBr3} nanocrystals}
\ce{CsPbBr3} NCs were synthesized according to a reported recipe \cite{Cao2020} with slight modifications. ODE (\SI{5}{mL}) and \ce{PbBr2} (\SI{69}{mg}) were mixed in a \SI{25}{mL} 3-neck flask and dried under vacuum for one hour at \SI{120}{\celsius}. Then, under Ar atmosphere, dried OA (\SI{0.5}{mL}) and dried OLA (\SI{0.5}{mL}) were injected to the mixture. The temperature was raised to \SI{180}{\celsius} and kept for \SI{10}{\minute}. Cs-oleate solution (\SI{0.4}{mL}) was swiftly injected, and after \SI{25}{\second} the reaction mixture was cooled by ice water bath.

For the purification of the NCs, the crude solution was centrifuged at \SI{6000}{rpm} for \SI{5}{\minute}. After the centrifuge, the supernatant was discarded and the particles were re-dispersed in anhydrous toluene forming colloidally stable solution.

The surface treatment of the colloidal \ce{CsPbBr3} NCs was performed following the procedure reported in reference\cite{Ahmed2018} with some modifications. Preparation of saturated \ce{NH4BF4} salt solution: toluene (\SI{2}{mL}, anhydrous) and \ce{NH4BF4} (\SI{10}{mg}) were stirred for \SI{10}{\minute}, sonicated for \SI{10}{\minute} and then centrifuged at \SI{6000}{rpm} for \SI{5}{\minute}. \ce{NH4BF4} salt precipitation was discarded, resulting in a saturated solution. \ce{NH4BF4} saturated solution (\SI{1}{mL}) was then stirred with \ce{CsPbBr3} NCs precipitation in toluene (\ce{0.25}{mL}) for \SI{30}{\minute}, creating surface treated \ce{CsPbBr3} NCs.

\subsection{Synthesis of \ce{CsPbI3} nanocrystals}
\ce{CsPbI3} NCs were synthesized according to reported in reference\cite{Pan2020} with minor modifications. ODE (\SI{5}{mL}), \ce{PbI2} (\SI{86.7}{mg}), OLA (\SI{1}{mL}, anhydrous), TDPA (\SI{120}{mg}) and TOPO (\SI{1.47}{mg}) were mixed in a \SI{50}{mL} 3-neck flask and dried under vacuum for one hour at \SI{120}{\celsius}. The temperature was raised to \SI{280}{\celsius} and kept for \SI{10}{\minute} under Ar atmosphere. Then Cs-oleate solution (\SI{0.4}{mL}) was quickly injected, and after \SI{15}{s} the reaction mixture was cooled by ice-water bath.

Purification procedure – crude solution was centrifuged at \SI{6000}{rpm} for \SI{5}{\minute}. Supernatant was discarded and precipitates were washed in anhydrous hexane, following additional centrifuge procedure (\SI{6000}{rpm} for \SI{5}{\minute}).

\section{Supporting analyses}
This section describes the additional analyses performed on the collected data, on-top of the heralded spectroscopy. It describes the estimation of the average number of absorbed photons per excitation pulse ($\ev{N}$), the zero delay normalized second order correlation of photon arrival times ($g^{(2)}(0)$) and the \emph{gated} zero delay normalized second order correlation of photon arrival times ($\Hat{g}^{(2)}(0)$).

\subsection{$\mathbf{\ev{N}}$}
The average number of absorbed photons per excitation pulse, $\ev{N}$, was estimated from the ratio of detected BX-1X photon pairs to the total number of single detections. This ratio can be defined as the following:
\begin{equation}\label{eqn:alpha}
\begin{split}
    \alpha \equiv \frac{N_2}{N_1} &= \frac{p_{abs}(\ge2) \cdot QY_{BX} \cdot QY_{1X} \cdot \eta \cdot {p_{det}}^2}{p_{abs}(\ge1) \cdot QY_{1X} \cdot \eta \cdot p_{det}}\\
    &= \frac{p_{abs}(\ge2)}{p_{abs}(\ge1)} \cdot QY_{BX} \cdot p_{det}
\end{split}
\end{equation}
$N_2$ is the number of detected photon pairs as described in the main text. $N_1$ is the number of single photons detected within a the 1X time-gate, i.e.\ during a time window of 0.5-\SI{30}{ns} following any excitation pulse (see \autoref{subsec:anaParams}). $p_{abs}(k)$ is the probability a NC absorbs $k$ photons in a single excitation pulse. $QY_{BX}$ and $QY_{1X}$ are the quantum yields of the BX and 1X, respectively. That is, the probability for the respective excited state to relax radiatively to a lower state. $\eta$ is a scalar factor accounting for single and pair detections filtered out due to the 1X temporal gate described above. $p_{det}$ is the probability to detect a photon that was emitted from the NC. Note that the temporal gating of $N_1$ serves not only to cancel out the factor of $\eta$ but also to filter out most contributions from the biexciton and trion states to the single-photon signal (see \autoref{sec:lifetime}).

$g^{(2)}(0)$, described in further detail in \autoref{subsec:g2}, is:
\begin{equation}\label{eqn:g2}
\begin{split}
    g^{(2)}(0) &= \frac{p_{abs}(\ge2) \cdot QY_{BX} \cdot QY_{1X} \cdot {p_{det}}^2}{\frac{{p_{abs}(\ge1)}^2}{2} \cdot {QY_{1X}}^2 \cdot {p_{det}}^2}\\
    &= \frac{2 \cdot p_{abs}(\ge2)}{{p_{abs}(\ge1)}^2} \cdot \frac{QY_{BX}}{QY_{1X}}
\end{split}
\end{equation}
In the first line, the nominator is the probability to absorb, emit and detect two photons following the same excitation pulse. The denominator represents the probability to absorb, emit and detect a single photon in each of two separate excitation pulses.

Absorption statistics are expected to follow a Poissonian distribution. That is, the probability to absorb $n$ photons in any single excitation pulse is:
\begin{equation}\label{eqn:p_abs}
    p_{abs}(n) = \frac{\ev{N}^n}{n!}\cdot e^{-\ev{N}}
\end{equation}
and hence:
\begin{equation}\label{eqn:p_abs>n}
    p_{abs}(\ge n) = 1-\sum_{k=o}^{n-1}\frac{\ev{N}^k}{k!}\cdot e^{-\ev{N}}
\end{equation}
Plugging this into \autoref{eqn:g2}, we see that for $\ev{N}\ll1$, the expression for $g^{(2)}(0)$ simplifies to the more commonly quoted expression: $g^{(2)}(0)\approx\frac{QY_{BX}}{QY_{1X}}$.

Finally, we can combine all the previous equations, to attain an expression for $\ev{N}$:
\begin{equation}\label{eqn:N}
    \ev{N} = -\ln\left(1 - \frac{2 \cdot \alpha}{QY_{1X} \cdot g^{(2)}(0) \cdot p_{det}} \right)
\end{equation}
$\alpha$ and $g^{(2)}(0)$ are measured quantities extracted from the same data used for the heralded spectroscopy. $QY_{1X}$ was measured, for an ensemble of NCs, by an absolute photoluminescnce (PL) quantum yield spectrometer (Quantaurus-QY, Hamamatsu), and is ${\sim}100\%$ for \ce{CsPbBr3} and ${\sim}42\%$ for \ce{CsPbI3}. $p_{det}$ was previously estimated for ${\sim}\SI{2.01}{eV}$ emission and a different grating\cite{Lubin2021} as $p_{det}\approx\num{1.5d-2}$. According to the factory characterization of the grating and the measured spectral response of the detector, we can estimate $p_{det}\approx\num{2.5d-2}$ for \ce{CsPbBr3} and $p_{det}\approx\num{1.2d-2}$ for \ce{CsPbI3}.

For the measurements shown in Figures 3 and 4 of the main text, $\ev{N}_{\ce{CsPbBr3}}=0.13\pm0.04$ and $\ev{N}_{\ce{CsPbI3}}=0.28\pm0.18$. For these $\ev{N}$ values, the probability to excite a NC more than twice is low ($p_{abs}(\ge3) \ll p_{abs}(\ge2)$). Combined with the typically lower quantum yields of higher multiexcitonic states (as evident in their shorter PL decay lifetimes\cite{DeJong2017}) we estimate that the contribution of triply and higher excited states to the heralded spectroscopy signal is negligible. We note that in both \autoref{eqn:alpha} and \autoref{eqn:g2} we neglect the contribution of higher multiexcitonic states. Some of these contributions cancel out in \autoref{eqn:N}, while others are negligible due to the low $\ev{N}$ and multiexciton quantum yield.

\subsection{$\mathbf{g^{(2)}(0)}$}
\label{subsec:g2}
\begin{figure}
    \centering
    \includegraphics[width=\linewidth]{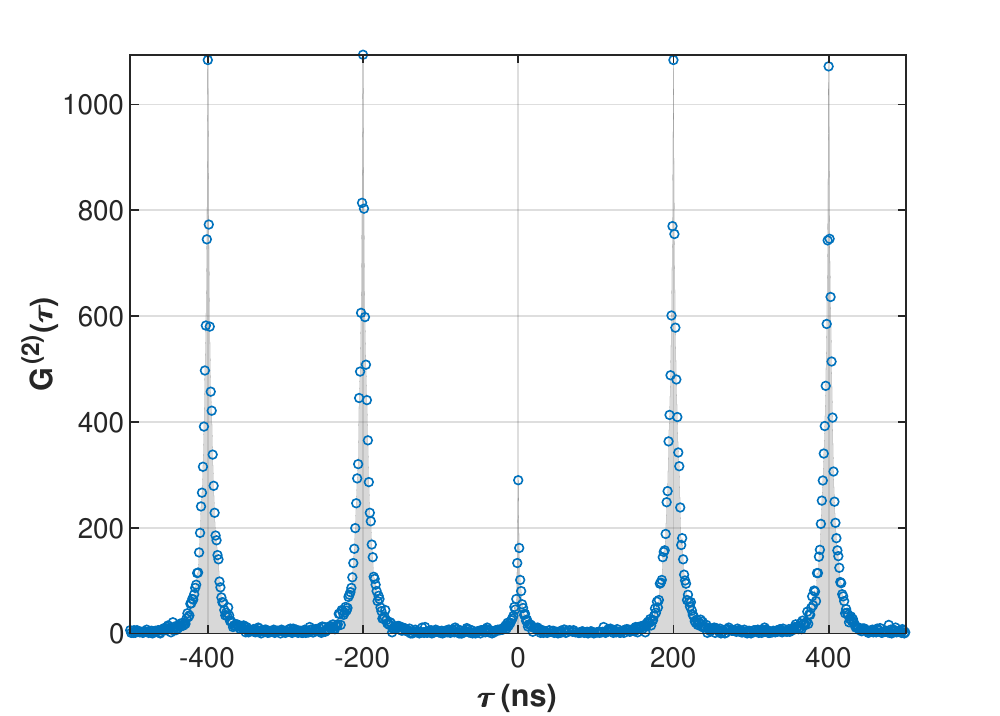}
    \caption{\textbf{Second order correlation of photon arrival time.} The second order correlation of photon arrival times for a single-NC measurement. The value of $G^{(2)}(\tau)$ represents the number of photon detection pairs with intra-detection delay of $\tau$ over the entire measurement. The attenuated zero delay peak is indicative of photon antibunching.}
    \label{fig:g2}
\end{figure}

$g^{(2)}(0)$ was calculated and corrected for errors arising from crosstalk and dark counts by the method described in reference\cite{Lubin2019}. Briefly, we treat the SPAD array pixels as the arms of a multiple-port Hanbury Brown and Twiss photon correlation setup. A histogram of detection pairs by the delay between the detections ($\tau$) is generated to extract the second order correlation of photon arrival times ($G^{(2)}(\tau)$). The number of detection pairs not originating in photon pairs (i.e.\ due to dark counts or inter-pixel crosstalk) is estimated from the measured intensity and SPAD array characterization, and subtracted from the histogram. \autoref{fig:g2} shows the corrected $G^{(2)}(\tau)$ extracted from the same single-NC measurement featured in Figure 2 of the main text. It features a series of peaks separated by the pulse repetition rate (\SI{200}{ns}), and widened due to the finite PL decay lifetime (${\sim}\SI{6}{ns}$, see \autoref{sec:lifetime}). The zero delay peak is visibly attenuated compared to the other peaks, indicating photon antibunching (a lower probability to detect two photons following the same excitation pulse as compared to detecting twice one photon following separate pulses). As described in the main text, this is due to the higher rate of non-radiative Auger processes in doubly-excited NCs, competing with radiative PL. The ratio between the area under the center peak, and the area under any other peak is termed the zero delay normalized second order correlation of photon arrival times, or $g^{(2)}(0)$. As described in the previous section, for the pump intesities used in this work, $g^{(2)}(0) \approx \frac{QY_{BX}}{QY_{1X}}$.

\subsection{$\mathbf{\Hat{g}^{(2)}(0)}$}
As described in the main text, $\Hat{g}^{(2)}(0)$ is estimated by first post-selecting only detections within 1 to \SI{30}{ns} after any excitation pulse, and then passing the filtered detections through the $g^{(2)}(0)$ analysis described in the previous subsection. The lower bound of the time-gate (\SI{1}{ns}) filters out multiexcitonic emission which features sub-ns PL decay lifetimes (see \autoref{sec:lifetime}). The upper bound (\SI{30}{ns}) serves to lower noise due to dark counts with minimal loss of 1X signal (as done in the heralded spectroscopy analysis, see \autoref{subsec:anaParams}). Due to the complexity of crosstalk correction in this case we omit the center $\pm \SI{625}{ps}$ of each $G^{(2)}(\tau)$ peak. This delay time window for the zero delay peak accounts for 99.5\% of crosstalk detection pairs.

\section{Photoluminescence decay lifetimes}
\label{sec:lifetime}
This section describes the methods used to estimate the 1X PL decay lifetime and an upper bound on the BX PL decay lifetime from the collected data. The results support the analysis parameter choices detailed in \autoref{subsec:anaParams}, and supply further reassurance to the identification of the 1X emission signal in heralded spectroscopy.

\subsection{Single-exciton}
\begin{figure}
    \centering
    \includegraphics[width=\linewidth]{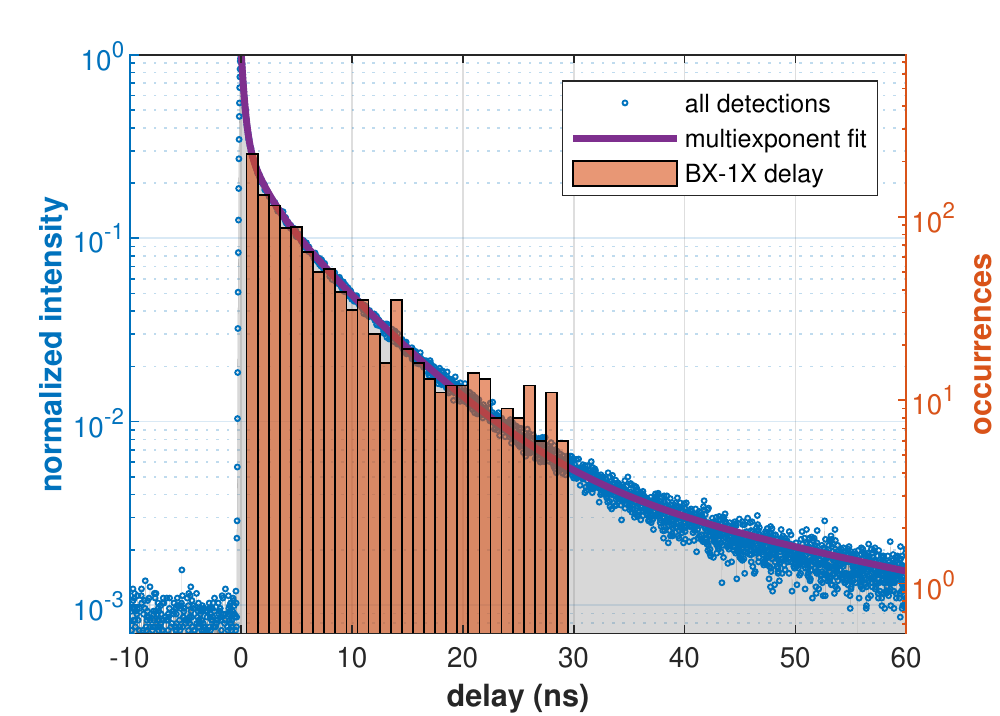}
    \caption{\textbf{Single-exciton photoluminescence decay lifetime.} Histogram of single-photon detection delays from the preceding excitation pulse, over a single-NC measurement (blue rings), and a fit to a multiexponential delay (purple line). Red bars are a histogram of delays between the two detections for all post-selected BX-1X pairs from the same measurement. To allow comparison, the right axis is scaled by a single scalar factor, such that the \SI{1}{ns} delay bins of both histograms coincide.}
    \label{fig:exLT}
\end{figure}

PL decay lifetime was estimated from a histogram of photon detections by their delay from the preceding excitation pulse. The blue trace in \autoref{fig:exLT} represents such a histogram for the single-NC measurement shown in Figure 2 of the main text. The purple trace represents a multiexponent fit of the form:
\begin{equation}
    y = \sum_k a_k \cdot e^{-\frac{t}{\tau_k}}
\end{equation}.
For this measurement the fitted coefficients were: $\tau_{1,2,3,4} \approx 0.3, 1.6 ,6.6, \SI{36.6}{ns}$ and $a_{1,2,3,4} \approx 0.69, 0.16, 0.20, 0.01$. The first two fast decay components have significant contribution only at the first ${\sim}\si{ns}$ following the excitation pulse. We estimate that they account for some combination of PL from multiexcitonic states, PL from the charged trion state\cite{Lubin2021} and the temporal instrument response function (IRF) of our system (see \autoref{subsec:IRF}). The long decay $\tau_4$ accounts for less than 1\% of the signal. Finally, $\tau_3 \approx \SI{6.6}{ns}$ is the dominant component between 0.5 and \SI{25}{ns} delay, and we assign it to the 1X PL decay lifetime. For the NCs in Figure 3 and 4 of the main text, 1X PL decay lifetimes are $5.9\pm\SI{1.6}{ns}$ (\ce{CsPbBr3}) and $8.1\pm\SI{1.4}{ns}$ (\ce{CsPbI3}).

Red bars in \autoref{fig:exLT} represent a histogram of the delay between BX and 1X detection for of all BX-1X photon pairs detected in the measurement by heralded spectroscopy. To allow comparison, the right axis is scaled only by a scalar factor compared to the left axis. The good temporal agreement between the intra-pair delay (red bars) and the 1X dominated 0.5-\SI{30}{ns} emission (blue trace) further supports the designation of the second photon of the pair as 1X emission, and $\tau_3\approx \SI{6.6}{ns}$ as its PL decay lifetime.

\subsection{Biexciton}
\begin{figure}
    \centering
    \includegraphics[width=\linewidth]{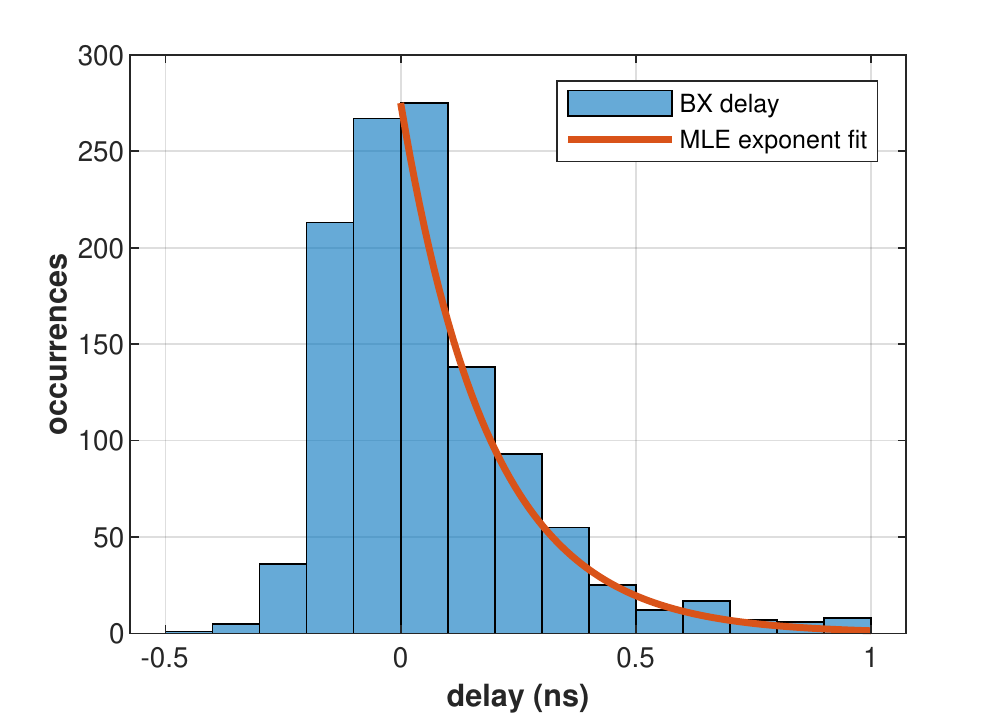}
    \caption{\textbf{Biexciton photoluminescence decay lifetime.} Histogram of BX detections' delay from the excitation pulse (blue bars). The observed temporal shape implies a convolution of the IRF (\autoref{fig:IRF}) and the BX PL decay lifetime. Red line represents a single-exponent fit using a maximum likelihood estimation on all detections with positive delay ($\tau \approx \SI{190}{ps}$).}
    \label{fig:bxLT}
\end{figure}

An upper bound for the BX PL decay lifetime is estimated from the delays between the first detections in each post-selected BX-1X pair and the preceding excitation pulse. \autoref{fig:bxLT} presents a histogram of such delays for the single-NC measurement featured in Figure 2 of the main text. Evidently, the distribution is a convolution of the IRF (\autoref{fig:IRF}) and the BX PL decay lifetime. To set an upper bound on the BX PL decay lifetime, we fit an single-exponent decay distribution to all positive BX delays, using a maximum likelihood estimate (red line, zero time delay is chosen as the delay with maximum single-photon detections). For this specific NC the result is $\tau \approx \SI{190}{ps}$. For the NCs featured in Figure 3 and 4 of the main text, the estimated upper bounds on BX PL decay lifetimes are $234\pm\SI{44}{ns}$ (\ce{CsPbBr3}) and $306\pm\SI{50}{ps}$ (\ce{CsPbI3}). Indeed, previously reported values lie within this bound\cite{Castaneda2016,DeJong2017,Ashner2019}.

The exact values of PL decay lifetime have no consequence for the validity of the heralded measurements, and are given here as an additional insight extracted from the same data-set. The approximate values and bounds, however, are used to justify the temporal-gating of BX and 1X detections in the heralded spectroscopy method (\autoref{subsec:anaParams}) and gated $\Hat{g}^{(2)}(0)$.

\section{System parameters}
The experimental apparatus and analysis parameters are detailed in reference\cite{Lubin2021}. The few modifications made to support the different PL parameters of the NCs used in this work are detailed in this section, and include: spectrometer grating, instrument response function (IRF), temporal gating values and number of array pixels used.

\subsection{Grating}
The grating used in this work is a 333 g/mm plane ruled reflection grating, with 5.7\textdegree\ nominal blaze angle (53-*-321R, Richardson). This resulted in a reciprocal linear dispersion of \num{2.8d-5} at the detector plane, and a spectral resolution of \SI{{\sim}4.5}{\angstrom}. The detector active pixel pitch is \SI{52.4}{\micro\meter}, and as a result, the pixel pitch in wavelength is \SI{{\sim}1.5}{nm}. This corresponds to pixel pitch of ${\sim}\SI{7}{meV}$ and ${\sim}\SI{4}{meV}$ at the emission spectral ranges of \ce{CsPbBr3} and \ce{CsPbI3}, respectively.

\subsection{Instrument response function}\label{subsec:IRF}
\begin{figure}
    \centering
    \includegraphics[width=\linewidth]{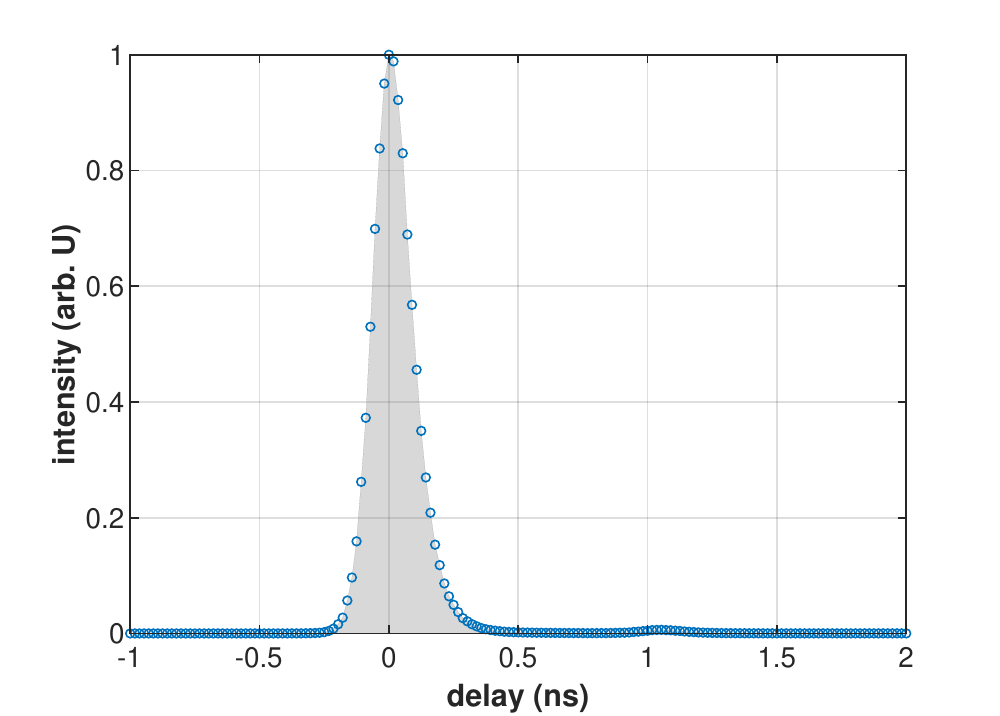}
    \caption{\textbf{Instrument response function.} The IRF, recorded by illuminating the SPAD array directly with the excitation laser. The presented histogram is generated according to the delay of each detection from the preceding excitation pulse, and summed 30 detector pixels. Zero time delay is chosen as the maximal intensity delay-bin.}
    \label{fig:IRF}
\end{figure}

The only update to the detector from reference\cite{Lubin2021} is an updated firmware that enables significantly better performance of the time-to-digital converters (TDC), and consequently an improved IRF. The IRF, seen in \autoref{fig:IRF}, is characterized by illuminating the detector directly with the synchronized excitation laser, and summing detections over 30 array pixels. The single-peak IRF features ${\sim}\SI{180}{ps}$ full width at half maximum (FWHM). This response is a convolution of the excitation pulse temporal width and the timing jitter of the pixels (${\sim}\SI{105}{ps}$ FWHM).

\subsection{Analysis parameters}
\label{subsec:anaParams}
Due to the improved IRF (see previous subsection), and shorter 1X fluorescence decay lifetime (see \autoref{sec:lifetime}), the temporal gates used to minimize the dark count rate (DCR) in reference\cite{Lubin2021} were refined: For the first photon of the pair (BX) only detections between \SIrange{-0.5}{1}{ns} delay from the fluorescence temporal peak were considered. Pairs were post-selected such that the second detection (1X) is detected within \SIrange{0.5}{30}{ns} following the first. Both BX and 1X upper gates are at least a factor of 3 longer than the respective fluorescence decay lifetime ($\tau$ in \autoref{sec:lifetime}), and thus serve to lower the DCR with negligible loss of signal. The lower bound of the BX (\SI{-0.5}{ns} from the fluorescence temporal peak) ensures detections before the overall fluorescence temporal peak are not lost. The lower bound of the 1X (\SI{0.5}{ns} after the BX) ensures correct identification of arrival order (it's significantly larger than the IRF FWHM), and filters out most of the inter-pixel crosstalk, as it is characterized by similar timescales as the IRF.

\subsection{Number of array pixels used}
To minimize DCR, only a subset of the detector's 512 pixels was utilized. In \ce{CsPbBr3} measurements, 30 pixels of the linear SPAD array were used, spanning the range of \SIrange{2.32}{2.53}{meV} photon energies. For \ce{CsPbI3} measurements, 43 pixels of the array were used, spanning the range of  \SIrange{1.76}{1.93}{meV} photon energies. One pixel in the 43 pixel range used for \ce{CsPbI3} was malfunctioning and was hence omitted from the presented analyses.

\section{Published values of biexciton binding energy in similar nanocrystals}

\autoref{table:BXBE} presents a list of previously reported values of the BX binding energy of cesium lead halide perovskite NCs. It includes only results for \ce{CsPbBr3} and \ce{CsPbI3} NCs, as investigated in this work. We adopt the convention used in the main text, where attractive exciton-exciton interaction is regarded as positive BX binding energy. Inspection of the data presented in the table reveals a lack of consensus among the reported values. In addition, an objective comparison of the measurements is made difficult by variations in the size and confinement regime of the particles studied. The last row of the table contains the ensemble results from our present work. 

\begin{table*}[t]
\centering
\begin{tabular}{ M{0.15\textwidth} M{0.15\textwidth} M{0.14\textwidth} M{0.12\textwidth} M{0.1\textwidth} M{0.12\textwidth} } 
\toprule

Reference & Technique\textsc{*} & Material & Edge length (nm) & $\ev{N}$  & BX binding energy (meV)\\ 

\midrule

% Wang's paper for values at low temps using power dependent PL. Not time resolved. 
Wang et al. (2015)\cite{} & %reference
PDPL (CRYO) & %technique
\ce{CsPbBr3} & %material
$9$ & %NC size
** &  %fluence
$\approx 50$ \\ %value
\midrule

% The Klimov paper. mostly about Iodide or mixed Br-I, which I exclude from this table.
Makarov et al. (2016)\cite{Makarov2016} & %reference
TA (SD) & %technique
\ce{CsPbI3} & %material
$11.2\pm 0.7$ & %NC size
$0.1$ & %<N> 
11 \\ %value
% % item
% & %reference
% \ce{CsPbBr_{1.5}I_{1.5}} & %material
% $10.7\pm 1.1 \text{ nm}$ & %NC size
%  & %technique
% ?? & %fluence
% 12  \\ %value
\midrule

% This is the paper with time resolved PL, whose results Tisdale and co. later attribute to sintering.
% Brimide values
Castaneda et al.\ (2016)\cite{Castaneda2016} & %reference
TRPL & %technique
\makecell{\ce{CsPbBr3} \\   \\ \ce{CsPBI3} \\ } & %material
\makecell{$\approx 7.4$ *** \\ $\approx 11.5$ ***  \\ $\approx 7.4$ ***  \\ $\approx 12.8$ *** } & %NC size
$\makecell{\approx 2}$ & %fluence
\makecell{$\approx 100 $ \\ $\approx 30 $\\ $\approx 90 $\\  $\approx 25 $} \\ %value
% Iodide values
%  & %reference
% \ce{CsPbI3} & %material
% V $\approx 100 \text{ nm}^3$ & %NC size
% & %technique
% $\approx 2$ & %fluence
% $\approx 40 $ \\ %value
\midrule

% ===================================================
% The paper from Angshuman and Adarsh.
Aneesh et al. (2017)\cite{Aneesh2017} & %reference
TA (SD) & %technique
\ce{CsPbBr3} & %material
$11$ & %NC size
$\approx 0.04$ & %fluence
$\approx 30$ \\ %value
\midrule

% ===================================================
% item
Yin et al. (2017)\cite{Yin2017} & %reference
SP (CRYO) & %technique
\ce{CsPbI3} & %material
$\approx 9$ & %NC size
$\approx 0.05$ & %fluence
$14.26 \pm 1.53$ \\ %value
\midrule

% ===================================================
% item
Yumoto et al. (2018) \cite{Yumoto2018} & %reference
TA (SD) & %technique
\ce{CsPbI3} & %material
$6$ & %NC size
$0.1$ & %fluence
$\approx 35$ \\ %value
\midrule

% ===================================================
% item
Ashner et al. (2019) \cite{Ashner2019} & %reference
TA & %technique
\ce{CsPbBr3} & %material
\makecell{$6$ \\ $8$ \\ $10$} & %NC size
$0.3$ & %fluence
\makecell{$-10$ \\ $-3$ \\ $-2$} \\ %value

% & %reference
% \ce{CsPbBr3} & %material
% $8$ & %NC size
%  & %technique
% $0.3$ & %fluence
% $-3$ \\ %value

%  & %reference
% \ce{CsPbBr3} & %material
% $10$ & %NC size
%  & %technique
% $0.3$ & %fluence
% $-2$ \\ %value
\midrule

% ===================================================
% item
Huang et al. (2021) \cite{Huang2020} & %reference
2DES (CRYO) & %technique
\ce{CsPbBr3} & %material
$9$ & %NC size
$< 0.1$ & %fluence
$\approx 100$ \\ %value
\midrule

% ===================================================
% item
Shen et al. (2021) \cite{Shen2021} & %reference
TA & %technique
\ce{CsPbBr3} & %material
$16$ & %NC size
\makecell{$6.42$ \\ $12.8$} & %fluence
\makecell{$61.2$ \\ $21.7$} \\ %value
% item
% & %reference
% \ce{CsPbBr3} & %material
% $16$ & %NC size
%  & %technique
% $12.8$ & %fluence
% $21.7$ \\ %value
\midrule

% ===================================================
% item
Dana et al. (2021) \cite{Dana2021} & %reference
TA (SD) & %technique
\ce{CsPbBr3} & %material
$6 \pm 0.7 $ & %NC size
$\ge 4$ & %fluence
$\sim -100$ \\ %value
\midrule

% ===================================================
% item
This work & %reference
HS & %technique
\makecell{\ce{CsPbBr3} \\ \ce{CsPbI3}} & %material
\makecell{$5.9\pm 1.3$ \\ $7.2\pm 1.9$} & %NC size
\makecell{$\approx0.1$ \\ $\approx0.3$} & %fluence
\makecell{$10\pm6$ \\ $1\pm9$} \\ %value

%  & %reference
% \ce{CsPbI3} & %material
% $7.2\pm 1.9$ & %NC size
% & %technique
% $\approx0.1$ & %fluence
% $1\pm9$ \\ %value
\bottomrule

\end{tabular}
\caption{\textbf{Measured values of the BX binding energy in cesium lead halide nanocrystals published in the literature.} $\ev{N}$ is the average number of photons absorbed per particle per pump pulse. Positive BX binding energy values correspond to an attractive exciton-exciton interaction.}
\label{table:BXBE}
\footnotesize{*\hspace{1pt} \textbf{PDPL} - Power dependent PL, \textbf{CRYO} - at cryogenic temperatures, \textbf{TA} - Transient absorption, \textbf{SD} - short delay, \textbf{TRPL} - Time resolved PL, \textbf{SP} - Single particle PL spectroscopy, \textbf{2DES} - Two-dimensional electron spectroscopy, \textbf{HS} - Heralded spectroscopy.}\\
\justifying{\footnotesize{**\hspace{8pt} No $\ev{N}$ quoted. Pump intensity varied from 4.5 to $\SI{54.7}{\mu J}$\hfill}}\\
\footnotesize{***\hspace{4pt}  Estimated from cross section data}.\\
\end{table*}

\cleardoublepage
\bibliography{supp.bib}